\documentclass[preprint2]{aastex}

\def \ie {i.e.,}

\shortauthors{Choi et al.}

\begin{document}


\title{Trigonometric Parallaxes of Star Forming Regions \\ 
       in the Perseus Spiral Arm}


\author{Y. K. Choi\altaffilmark{1,2}, K. Hachisuka\altaffilmark{3}, M. J. Reid\altaffilmark{4}, 
Y. Xu\altaffilmark{5}, A. Brunthaler\altaffilmark{1}, K. M. Menten\altaffilmark{1}, and T. M. Dame\altaffilmark{4}} 
\email{ykchoi@kasi.re.kr}


\altaffiltext{1}{Max-Planck-Institut f$\ddot{\textrm u}$r Radioastronomie, 
Auf dem H$\ddot{\textrm u}$gel 69, D-53121 Bonn, Germany}
\altaffiltext{2}{Korea Astronomy and Space Science Institute, 
776, Daedeokdae-ro, Yuseong-gu, Daejeon, 305-348, Korea}
\altaffiltext{3}{
Shanghai Astronomical Observatory, Chinese Academy of Science, Shanghai, 200030, China} 
\altaffiltext{4}{Harvard-Smithsonian Center for Astrophysics, 
60 Garden Street, Cambridge, MA 02138, USA}
\altaffiltext{5}{Purple Mountain Observatory, Chinese Academy of Sciences, 
Nanjing 210008, China}


\begin{abstract}
We report trigonometric parallaxes and proper motions of water masers 
for 12 massive star forming regions in the Perseus spiral arm 
of the Milky Way as part of the Bar and Spiral Structure Legacy (BeSSeL) Survey. 
Combining our results with 14 parallax measurements in the 
literature, we estimate a pitch angle of 9$^\circ$.9 $\pm$ 1$^\circ$.5 for 
a section of the Perseus arm.  The 3-dimensional Galactic peculiar motions 
of these sources indicate that on average they are moving toward the Galactic 
center and slower than the Galactic rotation.     
\end{abstract}

\keywords{astrometry --- Galaxy: kinematics and dynamics --- Galaxy: structure --- masers --- stars: formation}

\section{INTRODUCTION}

Although our Galaxy is known to be a barred spiral galaxy, 
it is very difficult to determine its structure owing to our location within its disk.
Considerable uncertainties still exist regarding
the number and locations of spiral arms, the length and
orientation of the central bar and the rotation curve.
Kinematic distances have often been used to infer spiral structure,  
however, they have large uncertainties stemming from
inaccuracies in the adopted Galactic rotation model and the existence of
significant peculiar (non-circular) motions \citep{xu06,reid09b}.  

Using very long baseline interferometry (VLBI), one can measure
trigonometric parallaxes to massive star forming regions with accuracies  
of order $\pm$10 $\mu$as \citep{honma12,reid13}.  
The Bar and Spiral Structure Legacy (BeSSeL) 
Survey\footnote{http://bessel.vlbi-astrometry.org/} 
is a National Radio Astronomy Observatory (NRAO)
\footnote{The National Radio Astronomy Observatory is a facility of 
the National Science Foundation operated under cooperative agreement 
by Associated Universities, Inc.}
key science project that aims to study the spiral structure and kinematics of our Galaxy 
by measuring trigonometric parallaxes and proper motions of hundreds of 
massive star forming regions with the Very Long Baseline Array (VLBA).  

The Perseus arm has been proposed as one of the two dominate spiral arms of the Galaxy, 
emanating from the far side of the Galactic bar \citep{Churchwell09}. 
Toward the Galactic anticenter, the Perseus arm is relatively close to the Sun and 
parallaxes and proper motions of 14 sources in the Perseus arm have 
already been obtained with the VLBA, the VLBI Exploration of 
Radio Astrometry (VERA) array, and the European VLBI Network (EVN)
\citep{asaki10, hachisuka09, Moellenbrock09, moscadelli09, niinuma11, oh10, 
reid09a, rygl10, sakai12, sato08, shiozaki11, xu06, zhang13}.

Here we present trigonometric parallaxes and proper motions of 
12 massive star forming regions in the Perseus spiral arm in the outer Galaxy. 
Combined with results from the literature (with one duplicate source, G094.60--1.79), 
we have a large sample of 25 sources located in the Perseus arm. 
These parallax measurements extend our sampling of the arm to smaller and 
larger Galactic longitudes (in the second and third quadrants) and essentially double 
the number of Perseus arm sources with trigonometric parallaxes. 
 
In Section 2, we describe our observations and data reduction. 
We present the parallax and proper motion results in Section 3. 
In Section 4, we discuss the Galactic locations and peculiar motions 
of the sources in the Perseus spiral arm. 
Finally, we summarize our results in Section 5.

\section{OBSERVATIONS}

\begin{deluxetable}{llcccccc}
\tabletypesize{\small}
\tablecaption{Observation Information \label{tbl-1}}
\tablewidth{0pt}
\tablehead{
\colhead{Project} & \colhead{Source} & \colhead{Epoch 1} & \colhead{Epoch 2} & \colhead{Epoch 3} & \colhead{Epoch 4} & \colhead{Epoch 5} & \colhead{Epoch 6}   } 
\startdata
BR145E & G100.37--3.57 &  2010 May 16  & 2010 Aug 09 & 2010 Oct 12 & 2010 Nov 27 & 2010 Dec 31& 2011 May 13 \\
\tableline
BR145F & G108.20+0.58  & 2010 Jun 05 & 2010 Sep 02 & 2010 Nov 26 & 2010 Dec 12 & 2011 Jan 30 & 2011 May 31 \\
       &   G108.47--2.81  & 2010 Jun 05 & 2010 Sep 02 & 2010 Nov 26 & 2010 Dec 12 & 2011 Jan 30 & 2011 May 31   \\
       &   G111.25--0.77    & 2010 Jun 05 & 2010 Sep 02 & 2010 Nov 26 & 2010 Dec 12 & 2011 Jan 30 & 2011 May 31       \\
\tableline
BR145G & G183.72--3.66 & 2010 Apr 23 & 2010 Jun 20 & 2010 Aug 23 & 2010 Sep 25 & 2010 Nov 16 & 2011 Mar 21 \\
\tableline
BR145K & G229.57+0.15  & 2010 Oct 07 & 2011 Jan 04 & 2011 Mar 09 & 2011 Apr 16 & 2011 Jun 02 & 2011 Oct 04 \\
       &    G236.81+1.98 & 2010 Oct 07 & 2011 Jan 04 & 2011 Mar 09 & 2011 Apr 16 & 2011 Jun 02 & 2011 Oct 04          \\       
       & G240.31+0.07 & 2010 Oct 07 & 2011 Jan 04 & 2011 Mar 09 & 2011 Apr 16 & 2011 Jun 02 & 2011 Oct 04     \\
\tableline
BR145P & G094.60--1.79 &  2011 May 15  & 2011 Aug 07 & 2011 Oct 21 & 2011 Nov 22 & 2012 Jan 08 & (2012 May 09) \\
       &  G095.29--0.93   &  2011 May 15  & 2011 Aug 07 & 2011 Oct 21 & 2011 Nov 22 & 2012 Jan 08 & 2012 May 09                    \\                
\tableline
BR145V & G108.59+0.49 & 2010 Dec 10 & 2011 Feb 21 & 2011 Apr 30 & 2011 May 29 & 2011 Jul 17 & 2011 Nov 25 \\
       &   G111.23--1.23   & (2010 Dec 10) & 2011 Feb 21 & 2011 Apr 30 & 2011 May 29 & 2011 Jul 17 & 2011 Nov 25     \\
\enddata
\tablecomments{Column 1 lists the project code. Each project has three target sources, 
but Column 2 only lists the massive star forming regions reported in this paper.
Each project was observed at 6 epochs, and the dates are shown in Columns 3 -- 8.
The dates in parentheses are observed, but not used for the parallax/proper motion fitting.
}
\end{deluxetable}

We performed multi-epoch observations with the VLBA under program BR145.
We observed the 6$_{16}$--5$_{23}$ masing transition of H$_2$O at a rest frequency of 
22.23508 GHz toward 12 star forming regions in outer Galaxy. 
Three maser sources projected close to each other on the sky were observed in one 
group, each with 2 to 4 background quasars. 
We observed each group at 6 epochs spread over one year. 
The dates of the observations are listed in Table \ref{tbl-1};  
they were optimized to allow an accurate parallax measurement for water
masers lasting seven months or longer.

In order to measure trigonometric parallaxes, we used phase-referenced observations, 
switching (every 20--30 s) between the maser target and an extragalactic continuum 
source, selected from previously known calibrators from ICRF1 \citep{ma98} 
and ICRF2\footnote{http://gemini.gsfc.nasa.gov/solutions/2010a/2010a.html} lists
or from our VLBA calibrator survey 
\citep{immer11}.  The observed sources are listed in Table \ref{tbl-2}.
Strong continuum sources were also observed to monitor delay and electronic phase 
differences among the recorded frequency bands.
In order to calibrate atmospheric delays, we placed four 0.5-hr ``geodetic blocks'' 
\citep{reid09a} spaced every two hours.

The data were correlated in two passes with the 
DiFX\footnote{This work made use of the Swinburne University of Technology software
correlator, developed as part of the Australian Major National Research
Facilities Programme and operated under license.} 
software correlator \citep{deller07} in Socorro, NM. 
Four dual-circularly polarized frequency bands of 8 MHz bandwidth were processed 
with 16 spectral channels for each frequency band. 
The one dual-polarized band that included the maser emission was re-processed with 
256 channels, giving a velocity channel spacing of 0.42 km s$^{-1}$. 
The data reduction was performed with the NRAO Astronomical Image Processing System 
(AIPS) package and ParselTongue scripts \citep{kettenis06} following the procedure 
described in \cite{reid09a}.

\begin{deluxetable}{lcccr}
\tabletypesize{\small}
\tablecaption{Source Information \label{tbl-2}}
\tablewidth{0pt}
\tablehead{
\colhead{Source} & \colhead{R.A. (J2000)} & \colhead{Decl. (J2000)} & \colhead{$\theta_{\rm sep}$} & \colhead{P.A.} \\
       & (h \ m \ s)  & ($^\circ$ \ ' \ ") & ($^\circ$) & ($^\circ$) 
}
\startdata
G094.60--1.79 & 21:39:58.2701 & +50:14:20.994  & --  & --  \\
J2137+5101   & 21:37:00.9862 & +51:01:36.129  & 0.9 & --31 \\
J2150+5103   & 21:50:14.2662 & +51:03:32.264  & 1.8 &  +63 \\
J2145+5147   & 21:45:07.6666 & +51:47:02.243  & 1.7 &  +28 \\
\tableline
G095.29--0.93 & 21:39:40.5089 & +51:20:32.808 & --  & --  \\
J2137+5101   & 21:37:00.9862 & +51:01:36.129 & 0.5 &  --127 \\
J2150+5103   & 21:50:14.2662 & +51:03:32.264 & 1.7 &  +100 \\
J2145+5147   & 21:45:07.6666 & +51:47:02.243 & 1.0 &  +62 \\
J2139+5300   & 21:39:53.6244 & +53:00:16.599 & 1.7 &  +1 \\
\tableline
G100.37-3.57 & 22:16:10.3651 & +52:21:34.113 & -- & -- \\
J2217+5202   & 22:17:54.4607 & +52:02:51.370 & 0.4 & +140 \\
J2209+5158   & 22:09:21.4869 & +51:58:01.833 & 1.1 & --111 \\
\tableline
G108.20+0.58 & 22:49:31.4775 & +59:55:42.006 & --  & --  \\
J2243+6055   & 22:43:00.8093 & +60:55:44.199 & 1.3 & --39 \\
J2254+6209   & 22:54:25.2930 & +62:09:38.725 & 2.3 & +15 \\
J2257+5720   & 22:57:22.0461 & +57:20:30.197 & 2.8 & +158 \\
\tableline
G108.47--2.81 & 23:02:32.0813 &+56:57:51.356 & --  & --  \\
J2301+5706   & 23:01:26.6266 &+57:06:25.508 & 0.2 & --46 \\
J2258+5719   & 22:58:57.9412 &+57:19:06.463 & 0.6 & --54 \\
J2257+5720   & 22:57:22.0461 &+57:20:30.197 & 0.8 & --62 \\
\tableline
G108.59+0.49 & 22:52:38.3150 & +60:00:51.888 & --  & --  \\
J2243+6055 & 22:43:00.8130 & +60:55:44.212 & 1.5 & --52  \\
J2254+6209 & 22:54:25.2930 & +62:09:38.725 & 2.2 & +6   \\
J2301+5706 & 23:01:26.6266 & +57:06:25.508 & 3.1 & +158  \\
J2258+5719 & 22:58:57.9412 & +57:19:06.463 & 2.8 & +163  \\
\tableline
G111.23-1.23 & 23:17:20.7888 & +59:28:46.970 & --  & --  \\
J2339+6010   & 23:39:21.1252 & +60:10:11.850 & 2.9 & +76  \\
J2258+5719   & 22:58:57.9412 & +57:19:06.463 & 3.2 & --132 \\
J2257+5720   & 22:57:22.0460 & +57:20:30.196 & 3.4 & --129 \\
J2254+6209   & 22:54:25.2926 & +62:09:38.724 & 3.9 & --46  \\
\tableline 
G111.25--0.77 & 23:16:10.3555 &+59:55:28.527 & --  & --  \\
J2339+6010   & 23:39:21.1251 &+60:10:11.850 & 2.9 & +85 \\
J2254+6209   & 22:54:25.2930 &+62:09:38.725 & 3.5 & --50 \\
J2258+5719   & 22:58:57.9412 &+57:19:06.463 & 3.4 & --139 \\
\tableline
G183.72-3.66 & 05:40:24.2276 & +23:50:54.728 & -- & -- \\
J0540+2507   & 05:40:14.3428 & +25:07:55.349 & 1.3 & --2 \\
J0550+2326  & 05:50:47.3909 & +23:26:48.177 & 2.4 & +100 \\
\tableline
G229.57+0.15  & 07:23:01.7718 & --14:41:34.339 & -- & -- \\
J0721--1530   & 07:21:13.4914 & --15:30:41.009 & 0.9 & --152 \\
J0724--1545   & 07:24:59.0063 & --15:45:29.370 & 1.2 & +156 \\
J0729--1320   & 07:29:17.8177 & --13:20:02.272 & 2.0 & +48 \\
J0721--1630   & 07:21:49.1377 & --16:30:19.746 & 1.8 & --171 \\
\tableline
G236.81+1.98  & 07:44:28.2367 & --20:08:30.606 & -- & -- \\
J0741--1937   & 07:41:52.7874 & --19:37:34.828 & 0.8 & --50 \\
J0745--1828   & 07:45:19.3291 & --18:28:24.799 & 1.7 & +7 \\
J0735--1735   & 07:35:45.8125 & --17:35:48.501 & 3.3 & --39 \\
J0739--2301   & 07:39:24.9981 & --23:01:31.885 & 3.1 & --158 \\
\tableline
G240.31+0.07  & 07:44:51.9676 & --24:07:42.372 & -- & -- \\
J0745--2451   & 07:45:10.2645 & --24:51:43.770 & 0.7 & +175 \\
J0749--2344   & 07:49:51.7793 & --23:44:48.788 & 1.2 & +72 \\
J0740--2444   & 07:40:14.7167 & --24:44:36.684 & 1.2 & --120 \\
\enddata
\tablecomments{Column 1 gives the names of the maser and background sources. 
Columns 2 and 3 list the absolute positions of the reference maser spot and background 
sources. Columns 4 and 5 give the angular separations ($\theta_{\rm sep}$) and 
position angles (P.A.)  east of north of the background sources relative to maser sources.}
\end{deluxetable}

\section{RESULTS}

In this section we present parallaxes and proper motions of 22 GHz H$_2$O masers 
for 12 massive star forming regions. 
The change in position of a maser spot relative to a background source is 
modeled with a sinusoidal parallax signature and a linear proper motion in 
right ascension and declination. 
Since systematic errors, associated with uncompensated atmospheric delays,
generally exceed those of random noise, we added ``error-floors'' in quadrature
to the formal position uncertainties in both coordinates. 
The error-floor values were determined by requiring the reduced $\chi_{\nu}^2$ 
of the post-fit residuals to be near unity in each coordinate \citep{reid09a}.   

\begin{figure*}
\epsscale{1.7}
\plotone{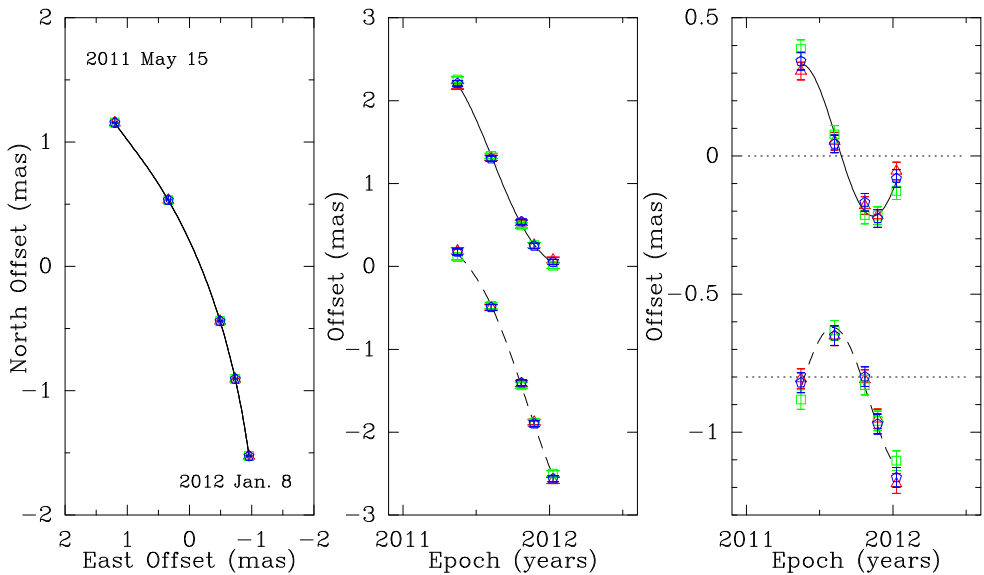}
\caption{Parallax and proper motion fit for G094.60--1.79.
Plotted are positions of the maser spot at $V_{\rm LSR}$ = --45.89 km s$^{-1}$ relative to three background sources 
J2137+5101 (red triangles), J2150+5103 (green squares) and J2145+5147 (blue hexagons), respectively.  
Left panel: sky-projected motion of the maser with respect to background source labeled with the first and last epochs. 
Middle panel: the position offsets of the maser along the east ($\alpha~\cos \delta$) and north direction ($\delta$) 
as a function of time. 
The best-fit model in the east and north directions are shown as continuous and dashed lines, respectively. 
Right panel: same as the middle panel but with fitted proper motions subtracted (parallax curve).   
Sixth epoch is not used for the fitting. 
(A color version of this figure is available in the online journal.)
\label{G094-para}}
\end{figure*}

\begin{figure*}
\epsscale{1.7}
\plotone{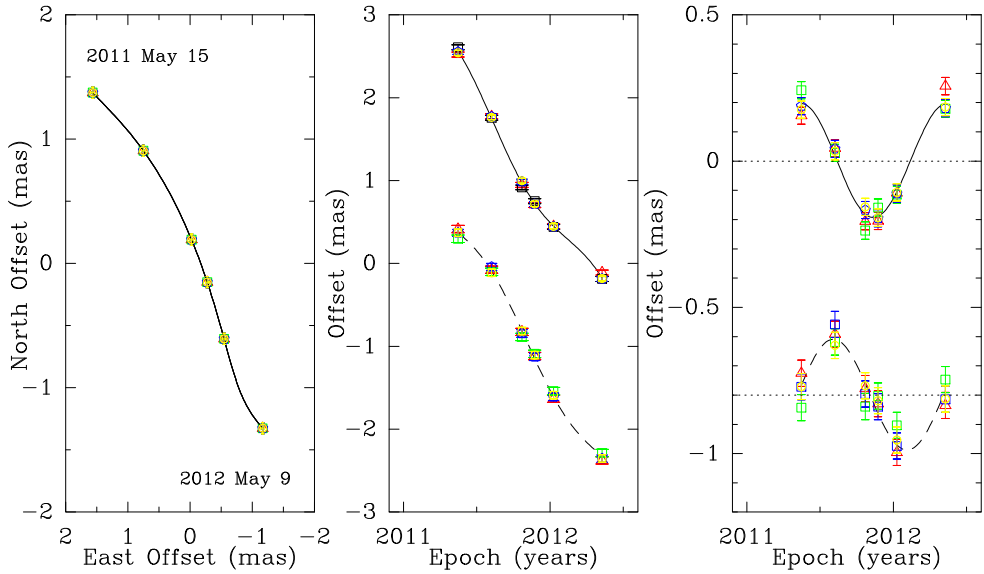}
\caption{Parallax and proper motion fit for G095.29--0.93. 
Plotted are positions of the maser spot at $V_{\rm LSR}$ = --35.67 km s$^{-1}$ relative to four background sources 
J2137+5101 (red triangles), J2150+5103 (green squares), J2145+5147 (blue hexagons) and J2139+5300 (yellow circle), respectively.  
The three panels are described in Figure \ref{G094-para}.
(A color version of this figure is available in the online journal.)
\label{G095-para}}
\end{figure*}

\begin{figure*}
\epsscale{1.7}
\plotone{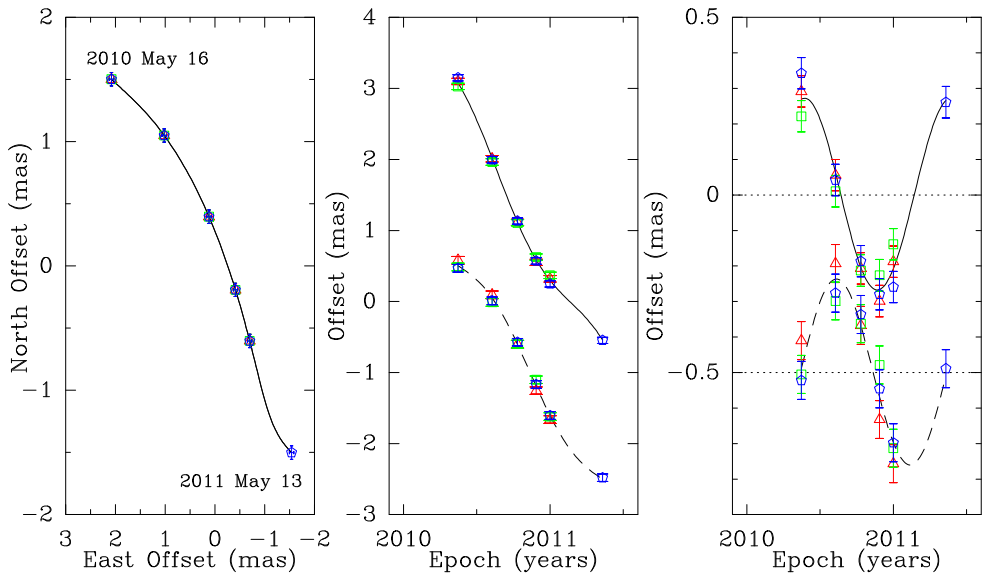}
\caption{Parallax and proper motion fitting for G100.37--3.57. 
Plotted are positions of the maser spot at $V_{\rm LSR}$ = --38.37 km s$^{-1}$ relative to two background sources 
J2209+5158 (red triangles), J2209+5158 (green squares) and J2217+5202 (blue hexagons), respectively.  
Since J2209+5158 has structures, the maser spot was detected in two separate regions. 
The three panels are described in Figure \ref{G094-para}.
(A color version of this figure is available in the online journal.)
\label{G100-para}}
\end{figure*}

\begin{figure*}
\epsscale{1.7}
\plotone{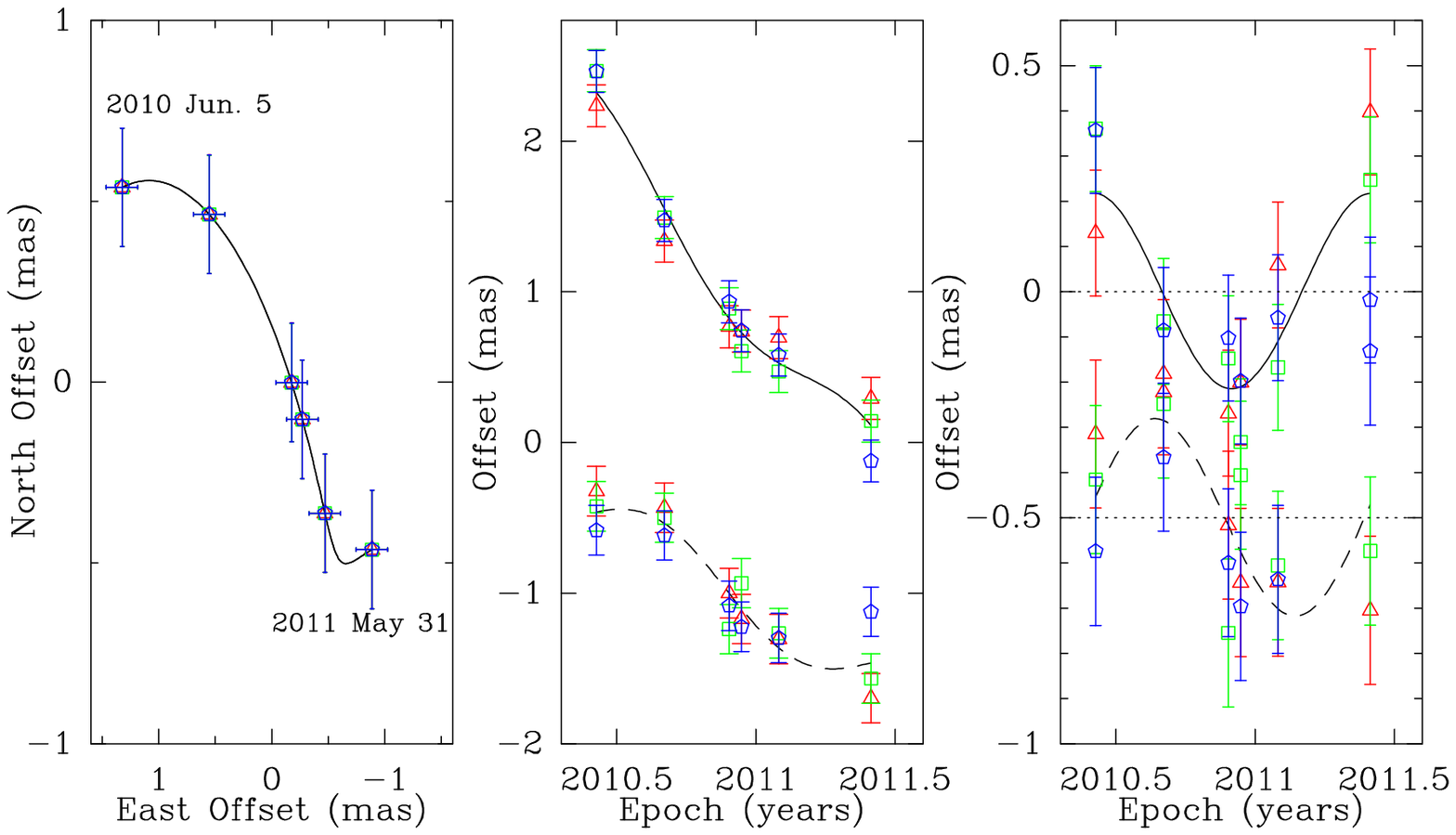}
\caption{Parallax and proper motion fit for G108.20+0.58. 
Plotted are positions of the maser spot at $V_{\rm LSR}$ = --54.16 km s$^{-1}$ relative to three background sources 
J2243+6055 (red triangles), J2254+6209 (green squares) and J2257+5720 (blue hexagons), respectively.  
The three panels are described in Figure \ref{G094-para}.
(A color version of this figure is available in the online journal.)
\label{G108A-para}}
\end{figure*}

\begin{figure*}
\epsscale{1.7}
\plotone{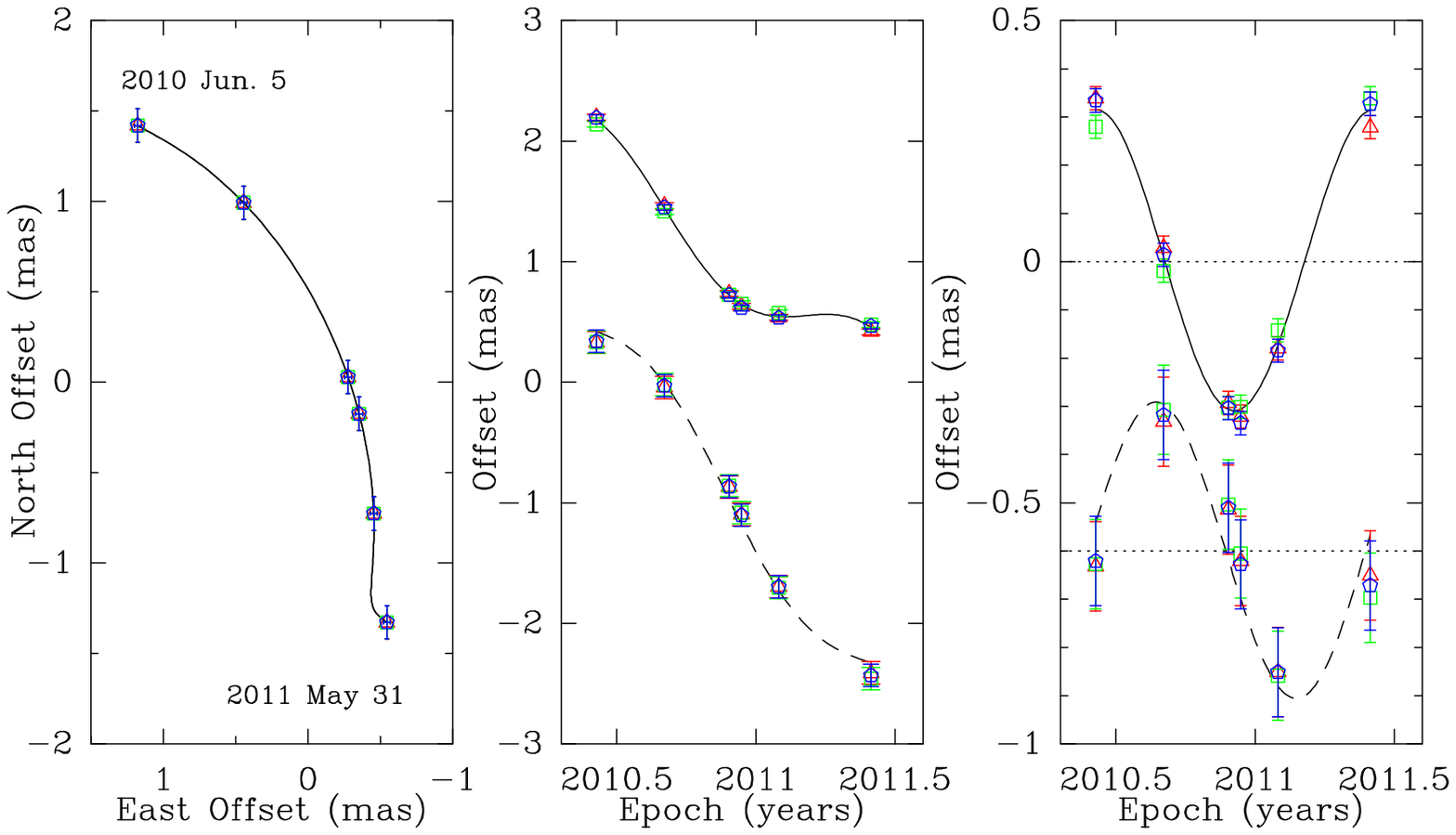}
\caption{Parallax and proper motion fit for G108.47--2.81.
Plotted are positions of the maser spot at $V_{\rm LSR}$ = --55.79 km s$^{-1}$ relative to three background sources 
J2301+5706 (red triangles), J2258+5719 (green squares) and J2257+5720 (blue hexagons), respectively.  
The three panels are described in Figure \ref{G094-para}.
(A color version of this figure is available in the online journal.)
\label{G108B-para}}
\end{figure*}

\begin{figure*}
\epsscale{1.7}
\plotone{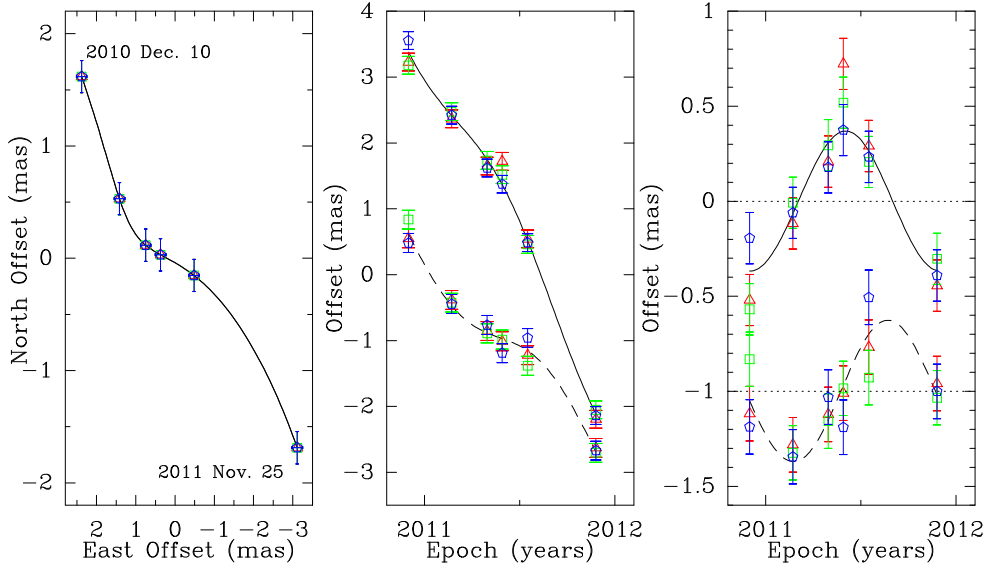}
\caption{Parallax and proper motion fit for G108.59+0.49. 
Plotted are positions of the maser spot at $V_{\rm LSR}$ = --53.47 km s$^{-1}$ relative to three background sources 
J2243+6055 (red triangles), J2254+6209 (green squares) and J2301+5706 (blue hexagons) and J2258+5719 (yellow circles), respectively.  
The three panels are described in Figure \ref{G094-para}.
(A color version of this figure is available in the online journal.)
\label{G108C-para}}
\end{figure*}

\begin{figure*}
\epsscale{1.7}
\plotone{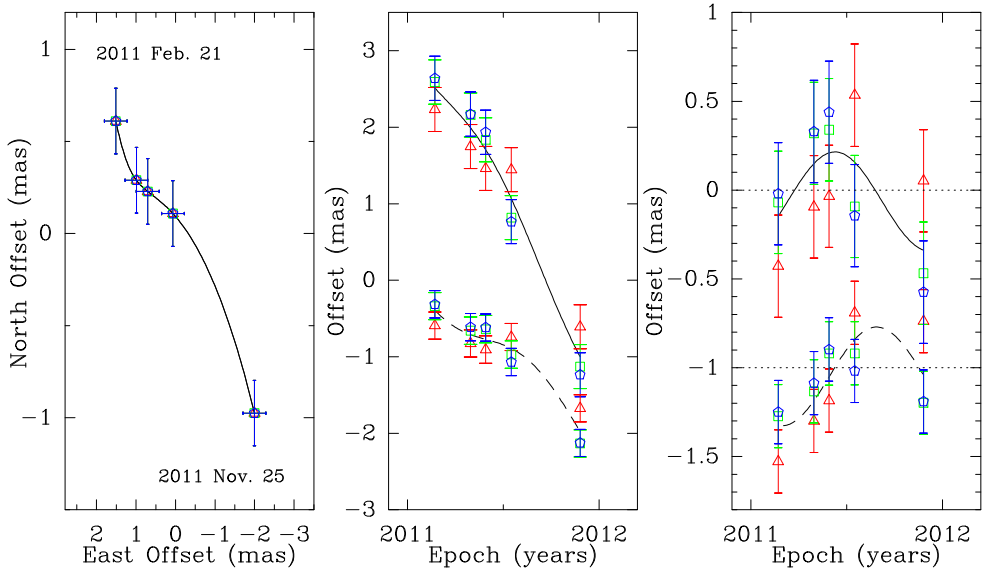}
\caption{Parallax and proper motion fit for G111.23--1.23. 
Plotted are positions of the maser spot at $V_{\rm LSR}$ = --49.94 km s$^{-1}$ relative to three background sources 
J2339+6010 (red triangles), J2258+5719 (green squares) and J2254+6209(blue hexagons), respectively.  
The three panels are described in Figure \ref{G094-para}.
1st epoch is not used for the fitting.
(A color version of this figure is available in the online journal.)
\label{G111A-para}}
\end{figure*}

\begin{figure*}
\epsscale{1.7}
\plotone{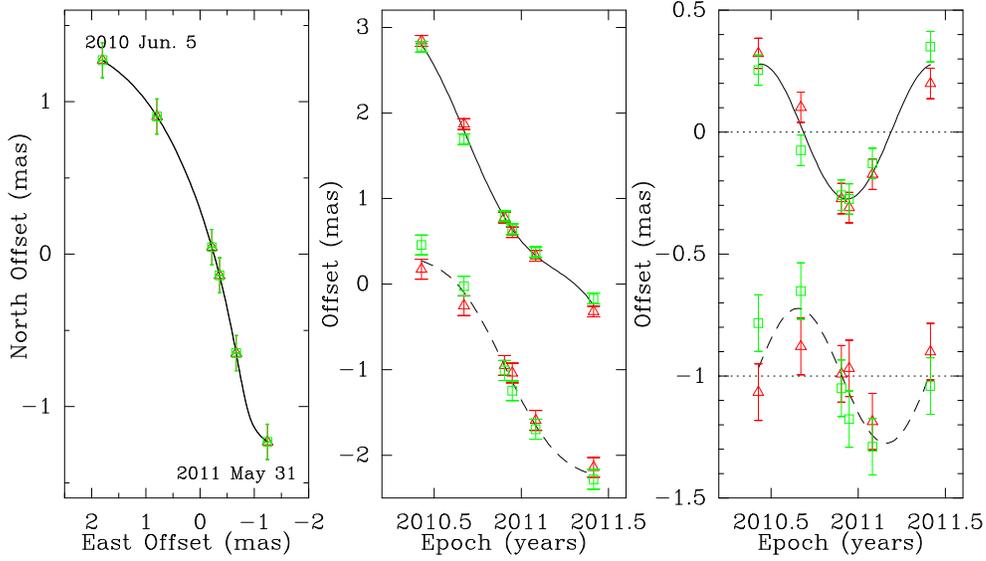}
\caption{Parallax and proper motion fit for G111.25--0.77.
Plotted are positions of the maser spot at $V_{\rm LSR}$ = 67.53 km s$^{-1}$ relative to two background sources 
J2339+6010 (red triangles) and J2258+5719 (green squares), respectively.  
The three panels are described in Figure \ref{G094-para}.
(A color version of this figure is available in the online journal.)
\label{G111B-para}}
\end{figure*}

\begin{figure*}
\epsscale{1.7}
\plotone{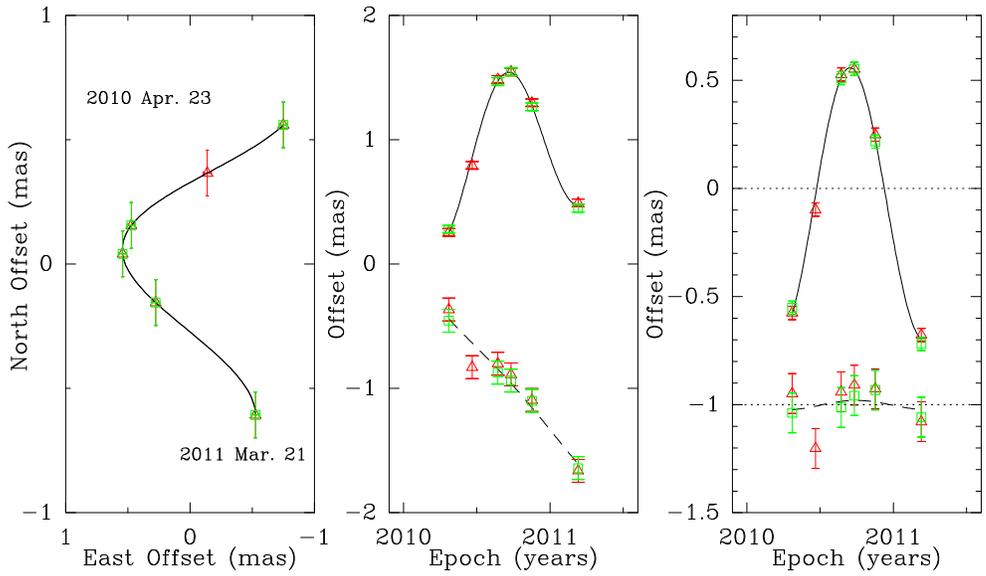}
\caption{Parallax and proper motion fit for G183.72--3.66. 
Plotted are positions of the maser spot at $V_{\rm LSR}$ = 4.58 km s$^{-1}$ relative to two background sources 
J0540+2507 (red triangles) and J0550+2326 (green squares), respectively.  
The three panels are described Figure \ref{G094-para}.
(A color version of this figure is available in the online journal.)
\label{G183-para}}
\end{figure*}

\begin{figure*}
\epsscale{1.7}
\plotone{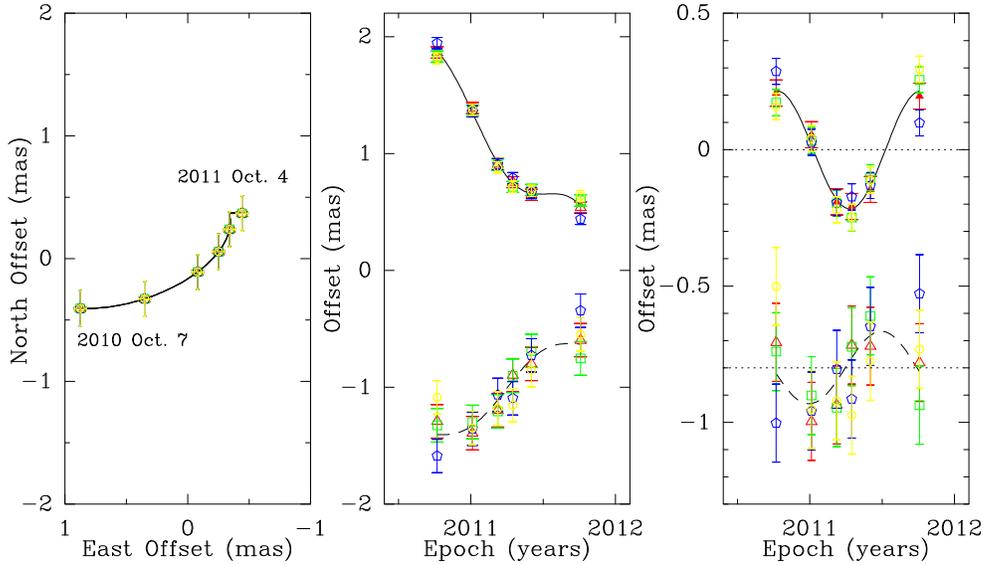}
\caption{Parallax and proper motion fit for G229.57+0.15. 
Plotted are positions of the maser spot at $V_{\rm LSR}$ = 57.11 km s$^{-1}$ relative to four background sources 
J0721--1530 (red triangles), J0724--1545 (green squares), J0729--1320 (blue hexagons) and J0721--1630 (yellow circles), respectively.  
The three panels are described in Figure \ref{G094-para}.
(A color version of this figure is available in the online journal.)
\label{G229-para}}
\end{figure*}

\begin{figure*}
\epsscale{1.7}
\plotone{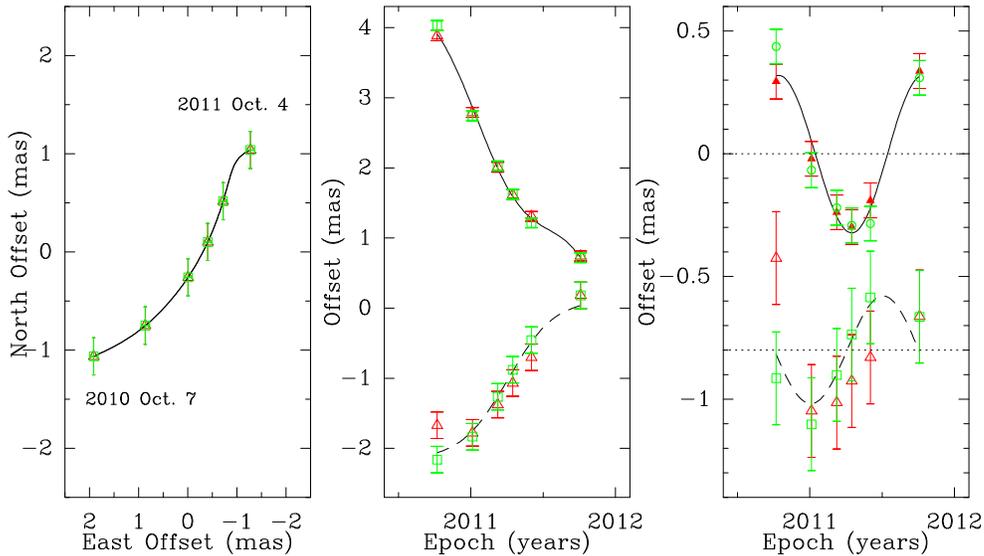}
\caption{Parallax and proper motion fit for G236.81+1.98. 
Plotted are positions of the maser spot at $V_{\rm LSR}$ = 42.31 km s$^{-1}$ relative to two background sources 
J0741--1937 (red triangles) and J0745--1828 (green squares), respectively.  
The three panels are described in Figure \ref{G094-para}.
(A color version of this figure is available in the online journal.)
\label{G236-para}}
\end{figure*}

\begin{figure*}
\epsscale{1.7}
\plotone{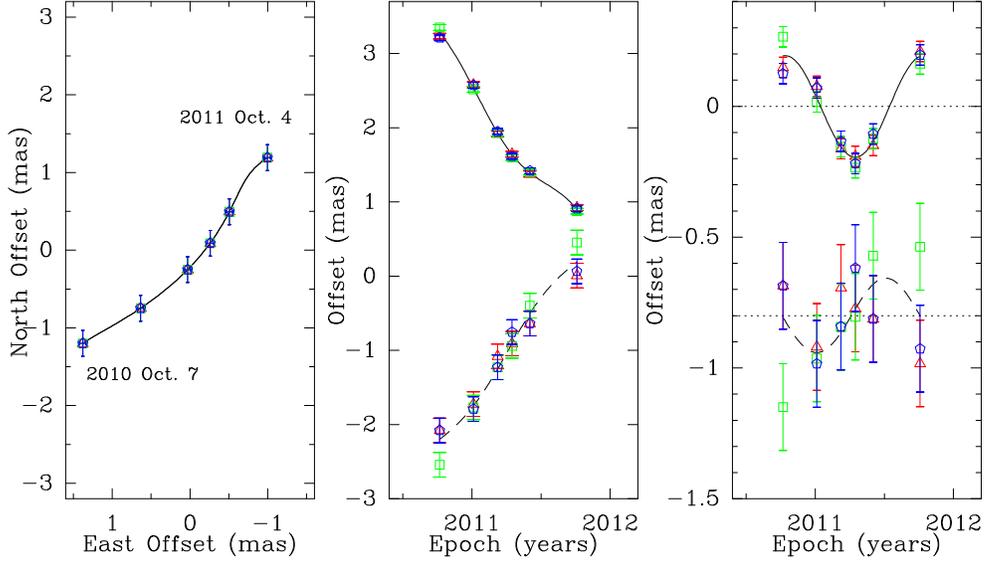}
\caption{Parallax and proper motion fit for G240.31+0.07.  
Plotted are positions of the maser spot at $V_{\rm LSR}$ = 67.53 km s$^{-1}$ relative to three background sources 
J0745--2451 (red triangles), J0749--2344 (green squares), J0740-2444 (blue hexagons), respectively.  
The three panels are described in Figure \ref{G094-para}. 
(A color version of this figure is available in the online journal.)
\label{G240-para}}
\end{figure*}

A bright H$_2$O maser spot was used as the phase reference in order to calibrate
the continuum source data and then measure position offsets used for
parallax and proper motion fitting. 
When we estimated a parallax using several maser spots in one source, 
the quoted parallax uncertainty is the formal fitting uncertainty multiplied by 
$\sqrt{N}$ (where $N$ is the number of maser spots) in order to account for 
possible correlations among the position measurements for the maser spots. 
Figures \ref{G094-para}--\ref{G240-para} show positions for the maser spots 
relative to the background sources as a function of time
and the parallax and proper motions fits. 
The fitting results are summarized in Table \ref{tab-para}.
Detailed information for each source is presented in the Appendix. 

\begin{deluxetable}{lccrrr}
\tablecaption{Parallaxes and Proper Motions \label{tab-para}}
\tablewidth{0pt}
\tablehead{
\colhead{Source} & \colhead{Parallax} & \colhead{Distance} & \colhead{$\mu_x$} & \colhead{$\mu_y$} & \colhead{$V_{\rm LSR}$}\\
                & (mas) & (kpc)  & (mas yr$^{-1}$) & (mas yr$^{-1}$)  & (km s$^{-1}$)  
}
\startdata
G094.60--1.79 & 0.253 $\pm$ 0.024 & 3.95 $^{+0.41}_{-0.34}$ & --2.59 $\pm$ 0.24 & --4.02 $\pm$ 0.34 & --49 $\pm$ \ 5  \\ 
G095.29--0.93 & 0.206 $\pm$ 0.007 & 4.85 $^{+0.17}_{-0.16}$ & --2.75 $\pm$ 0.20 & --2.76 $\pm$ 0.20 &  --38 $\pm$ \ 5 \\ 
G100.37--3.57 & 0.289 $\pm$ 0.016 & 3.46 $^{+0.20}_{-0.18}$ & --3.66 $\pm$ 0.60 & --3.02 $\pm$ 0.60 & --37 $\pm$ 10 \\
G108.20+0.58 & 0.227 $\pm$ 0.037 & 4.41 $^{+0.86}_{-0.62}$ & --2.25 $\pm$ 0.50 & --1.00 $\pm$ 0.50 &  --49 $\pm$ \ 5 \\
G108.47--2.81 & 0.309 $\pm$ 0.010 & 3.24 $^{+0.11}_{-0.10}$ & --3.13 $\pm$ 0.49 & --2.79 $\pm$ 0.46 & --54 $\pm$ \ 5 \\
G108.59+0.49 & 0.405 $\pm$ 0.033 & 2.47 $^{+0.22}_{-0.19}$ & --5.56 $\pm$ 0.40 & --3.40 $\pm$ 0.40 & --52 $\pm$ \ 5 \\
G111.23--1.23 & 0.300 $\pm$ 0.081 & 3.33 $^{+1.23}_{-0.71}$ & --4.37 $\pm$ 0.60 & --2.38 $\pm$ 0.60 & --53 $\pm$ 10 \\
G111.25--0.77 & 0.299 $\pm$ 0.022 & 3.34 $^{+0.27}_{-0.23}$ & --2.02 $\pm$ 0.52 & --2.31 $\pm$ 0.52 & --43 $\pm$ \ 5 \\
G183.72--3.66 & 0.629 $\pm$ 0.012 & 1.59 $^{+0.03}_{-0.03}$ & +0.38 $\pm$ 1.30 & --1.32 $\pm$ 1.30 &  +3 $\pm$ \ 5 \\
G229.57+0.15 & 0.218 $\pm$ 0.012 & 4.59 $^{+0.27}_{-0.24}$ & --1.33 $\pm$ 0.70 & +0.77 $\pm$ 0.70 & +47 $\pm$ 10 \\
G236.81+1.98 & 0.326 $\pm$ 0.026 & 3.07 $^{+0.27}_{-0.23}$ & --2.49 $\pm$ 0.43 & +2.67 $\pm$ 0.41 &  +43 $\pm$ \ 7\\
G240.31+0.07 & 0.188 $\pm$ 0.016 & 5.32 $^{+0.49}_{-0.42}$ & --2.43 $\pm$ 0.20 & +2.49 $\pm$ 0.20 & +67 $\pm$ \ 5 \\
\enddata
\tablecomments{Column 1 lists source names. Columns 2 and 3 give the measured parallax and the  
distance converted from the parallax. 
Columns 4 and 5 are proper motions in the eastward
($\mu_x$=$\mu_{\alpha} cos\delta$) and northward ($\mu_y$=$\mu_{\delta}$) directions.
Column 6 lists $V_{\rm LSR}$ of the star forming region.}  
\end{deluxetable}

When there were numerous maser spots, we attempted to fit an expanding model to the internal 
(relative) motions (as in G108.47--2.81, G111.25--0.77, and G236.81+1.98) in order to
estimate the motion of the central, exciting star directly. 
Details of this model fitting procedure are in \cite{sato10}. 

For sources with few maser spots, the proper motion for the central star 
was estimated from an unweighted average of all the measured absolute proper motions.
For these cases, since water maser motions relative to the central star typically
are tens of km s$^{-1}$, we adopted a 5 to 15 km s$^{-1}$ uncertainty for each velocity 
component, converted to an angular motion with the measured distance.
Several criteria were used to estimate the uncertainties in the motion components of the central star, 
depending on the complexity and width of the maser spectrum (as an indication of the likely outflow speed) 
and the difference between the water maser $V_{\rm LSR}$ values and the thermal CO value
(also an indication of the magnitude and likelihood of un-modeled outflow issues).
While $\pm5$ km s$^{-1}$ was used as the additional uncertainty for the components 
of  motion for G095.29--0.93, G108.59+0.49, and G240.31+0.07, 
$\pm10$ km s$^{-1}$ was used for G094.60--1.79, G100.37--3.57, G108.20+0.58, G111.23--1.23, and G183.72--3.66, 
 and $\pm15$ km s$^{-1}$ was used for G229.57+0.15.

\section{The Perseus Arm}

\subsection{Location and Pitch Angle}

\begin{figure*}
\includegraphics[scale=0.9]{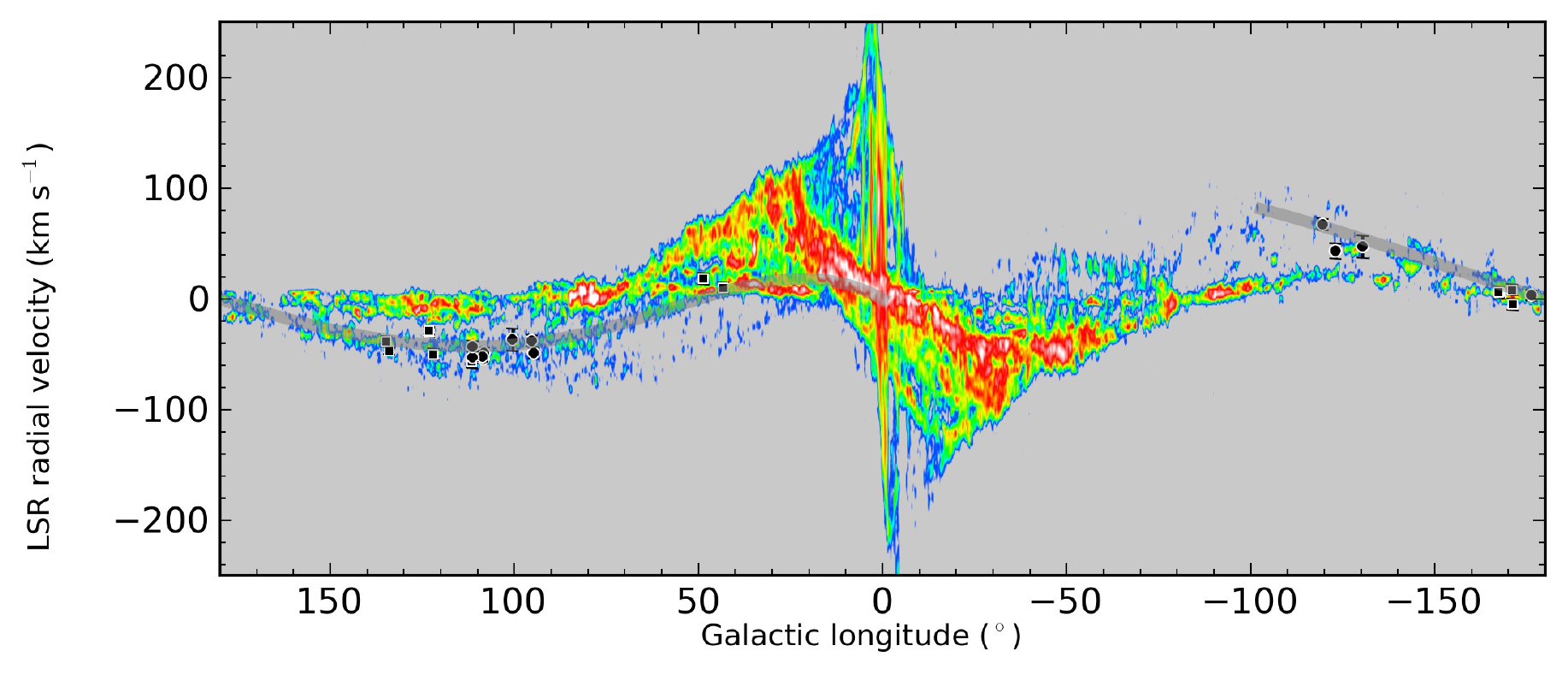}
\caption{Locations of high mass star forming regions with parallaxes
superposed on a CO $l-v$ diagram from the CfA 1.2-m survey  \citep{dame01}.
Sources presented in this paper are indicated with circles and those from 
literature with squares. The grey shadowed line traces the Perseus spiral 
arm \citep{vallee08}; the width of the line denotes a $\pm 10$ km s$^{-1}$ 
velocity dispersion.      
(A color version of this figure is available in the online journal.)
\label{lv}}
\end{figure*}

The maser sources were assigned to the Perseus arm based on matching
their longitudes and velocities to a prominent ``track'' in the longitude--velocity 
($l-v$) space of CO emission that is generally associated with this arm.
Note, that we did not use our parallax measurements to assign sources to the arm.  
This removes a potential selection bias for defining the arm and its 
properties. We plot these sources on a CO $l-v$ diagram from \cite{dame01} 
in Figure \ref{lv}. 

\begin{figure*}
\includegraphics[angle=-90,scale=0.8]{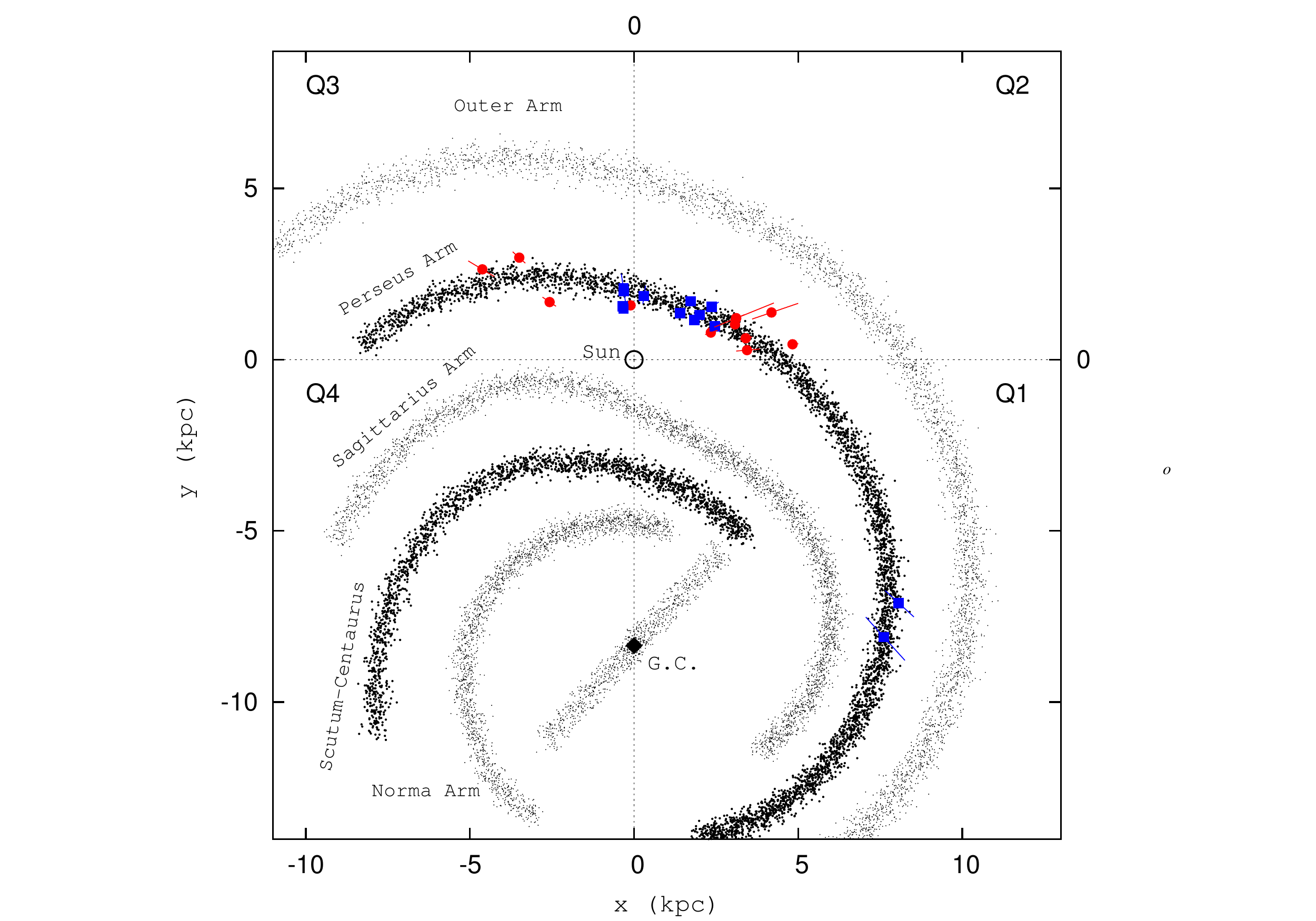}
\caption{Schematic view of the spiral arms of the Milky Way after \cite{taylor93} with updates. 
$R_0$ = 8.34 kpc \citep{reid13} is assumed. 
The location of the central bar \citep{Benjamin05} is also reported. 
The positions of the Perseus arm sources previously measured with trigonometric parallaxes are shown in blue squares 
together with the present measurements in red circles.   
(A color version of this figure is available in the online journal.)
\label{Gal-loc}}
\end{figure*}

With trigonometric parallax measurements, we can accurately locate the massive star 
forming regions of the Perseus arm of the Milky Way. 
Assuming the distance to the Galactic center $R_0$ to be 8.34 kpc \citep{reid13},
Figure \ref{Gal-loc} shows the locations of our sources in the Milky Way as red circles, 
together with previous results from the literature as blue squares.

For simplicity, we assume that a section of a spiral arm follows a log-periodic function: 
$$\ln{R} = \ln{R_{ref}} - (\beta - \beta_{ref}) \tan\psi~~~\eqno{(1)},$$ 
where $R$ and $\beta$ are Galactocentric radius and azimuth, respectively,
$R_{ref}$ is the radius at a reference azimuth $\beta_{ref}$, and $\psi$ is
the spiral pitch angle (\ie\ the angle between a spiral arm and a tangent to a 
Galactocentric circle).  Galactocentric azimuth is defined as the angle between the 
Sun and the source as viewed from the center, with azimuth increasing with Galactic 
longitude in the first quadrant.  We estimated two parameters, $R_{ref}$ and $\psi$
using a Bayesian approach, which minimized
the ``distance'' from the straight line described by Equation ~(1), variance weighted by the
uncertainty in this direction, and described in detail in \citet{reid13}.

\begin{figure*}
\includegraphics[angle=0,scale=1.3]{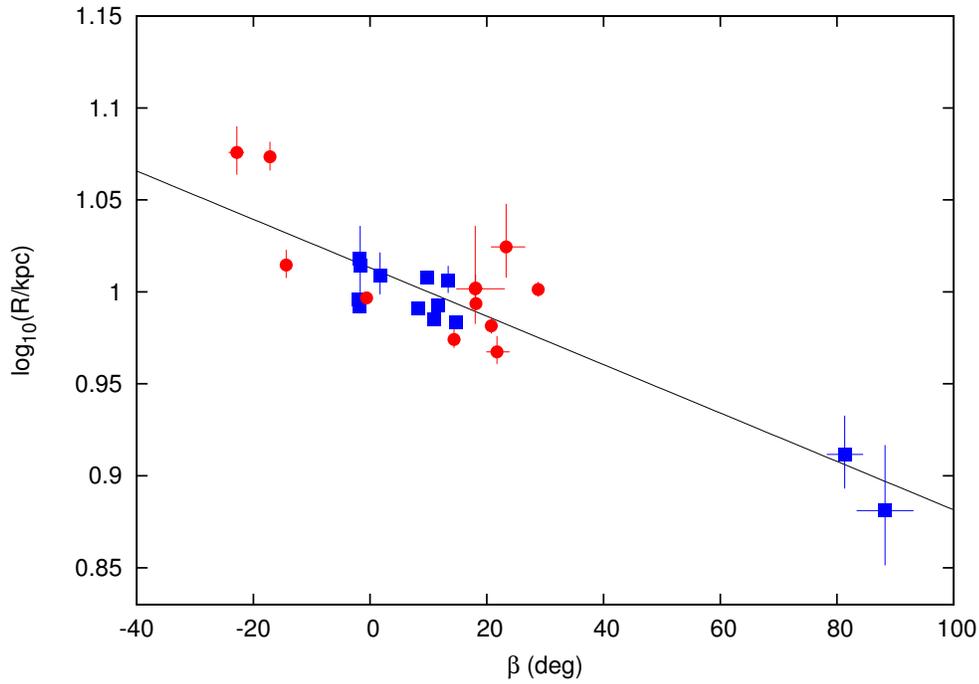}
\caption{Plot of the logarithm of Galactocentric radius (R) vs. Galactocentric azimuth ($\beta$) 
for the Perseus spiral arm sources with our measurements in red circles and the literature in blue squares.
The slope of the line is proportional to tangent of a pitch angle, $\psi$.   
(A color version of this figure is available in the online journal.)
\label{pitch}}
\end{figure*}

Using the 25 sources with parallax measurement that are assigned to the Perseus arm 
(based on matching to CO $l-v$ tracks), we estimate the arm pitch angle  
to be 9$^\circ$.9 $\pm$ 1$^\circ$.5, with an arm width of 0.38 kpc,
where $\ln{R_{ref}}$ = 2.29 $\pm$ 0.01 at $\beta_{ref}$ = 13$^\circ$.6
(uncertainties give 68\% confidence ranges).
The arm width was estimated by adding an ``astrophysical noise'' term,
reflecting a non-zero arm width, in quadrature with measurement uncertainty 
when fitting for the pitch angle.  The magnitude of this noise term was adjusted
to give a $\chi_\nu^2$ per degree of freedom of unity for the residuals.
In Figure \ref{pitch}, we plot the data and fitted line whose slope is tan$\psi$. 

Pitch angles for the Perseus spiral arm have previously been estimated by 
\cite{reid09b} and \cite{sakai12}.  \cite{reid09b} reported a value of 
$16^\circ.5 \pm 3^\circ.1$ based on data from 4 sources and 
\cite{sakai12} reported $17^\circ.8 \pm 1^\circ.7$ from 8 sources.
Both used a least-squares fitting approach that minimized the residuals
in the ``y-axis'' ($\ln{R}$) only.  Since distance uncertainties affect both
axes in Equation ~(1), \ie\ $\ln{R}$ and $\beta$, a better approach is to 
minimize the residuals {\it perpendicular} to the best fit line.
As a test, we refit the data used by \cite{reid09b} with our
improved Bayesian approach and obtained an estimated pitch angle of 
$15^\circ.1 \pm 6^\circ.4$.  This gives a similar pitch angle as the previous estimate, 
but with a larger and more realistic uncertainty.  With this uncertainty, the old 
result is statistically consistent with the new result.  Of course the new result 
benefits from using a much larger sample and should be preferred.

Based on the spiral pattern fit results, we obtain a distance between the Sun and the Perseus spiral arm
of 2.0 $\pm$ 0.2 kpc in the anticenter direction ($l$ = 180$^\circ$, $\beta$ = 0$^\circ$). 
Since the Sagittarius arm is located at 1.4 $\pm$ 0.2 kpc from the Sun \citep{yuanwei}, 
the Perseus arm is more distant from the Sun than the Sagittarius arm.

\subsection{Peculiar Motions}

Using the parallax distances to convert proper motions to linear speeds and
combining with Doppler radial velocities, 
we can calculate the three-dimensional space motions for the massive star forming 
regions in the Perseus spiral arm.  
Assuming $R_0$ = 8.33 $\pm$ 0.16 kpc, $\Theta_0$ = 243 $\pm$ 6 km s$^{-1}$, 
dT/dR = --0.2 $\pm$ 0.4 km s$^{-1}$ kpc$^{-1}$  
and Solar Motion components of $U_{\odot}$ = 10.7 $\pm$ 1.8 km s$^{-1}$, 
$V_{\odot}$ = 12.2 $\pm$ 2.0 km s$^{-1}$ and 
$W_{\odot}$ = 8.7 $\pm$ 0.9 km s$^{-1}$ from B1 model in \cite{reid13}
which used as a strong prior the \citet{schoenrich10} Solar Motion,
we calculate the peculiar motion components (U$_s$, V$_s$, W$_s$) of the sources.  
U$_s$ is toward the Galactic center, 
V$_s$ is in direction of Galactic rotation, and 
W$_s$ is toward the north Galactic pole.
The estimated peculiar motions for the sources are listed in Table \ref{tab-uvw}.

The variance-weighted average of the peculiar motion components
for 25 sources in the Perseus arm are 
$\overline{U_s} =  9.2 \pm 1.2$ km s$^{-1}$ toward the Galactic center,
$\overline{V_s} = -8.0 \pm 1.3$ km s$^{-1}$ in the direction of Galactic rotation, and 
$\overline{W_s} = -2.3 \pm 1.1$ km s$^{-1}$ toward the north Galactic pole.  
For the comparison, when we assume 
$R_0$ = 8.34 $\pm$ 0.16 kpc, $\Theta_0$ = 240 $\pm$ 8 km s$^{-1}$, 
dT/dR = --0.2 $\pm$ 0.4 km s$^{-1}$ kpc$^{-1}$  
and Solar Motion components of $U_{\odot}$ = 10.7 $\pm$ 1.8 km s$^{-1}$, 
$V_{\odot}$ = 15.6 $\pm$ 6.8 km s$^{-1}$ and 
$W_{\odot}$ = 8.9 $\pm$ 0.9 km s$^{-1}$ from A5 model in \cite{reid13}, 
we obtained 
$\overline{U_s} = 9.4 \pm 1.3 $ km s$^{-1}$,
$\overline{V_s} = -4.4 \pm 2.0 $ km s$^{-1}$, and 
$\overline{W_s} = -2.1 \pm 1.1 $ km s$^{-1}$.
The average peculiar motion in the direction of Galactic rotation is affected by the $V_{\odot}$ component of solar motion.

These peculiar motion values indicate that on average sources in the Perseus arm are moving toward 
the Galactic center and move slower than for circular Galactic orbits.
Compared with the average peculiar motions of the 100 sources with parallaxes in many arms, 
($\overline{U_s} = 2.9 \pm 2.1 $ km s$^{-1}$ and $\overline{V_s} = -5.0 \pm 2.1 $ km s$^{-1}$ from B1 model, \citet{reid13}), 
the Perseus arm sources have higher average peculiar motions and it seems to be a characteristic of the arm at least in the second and third quadrants.   

The general claim of the spiral density wave theory in literature from 1960s is 
that peculiar motions should be toward the Galactic center and counter to the Galactic rotation. 
Our result is consistent with that claim. 

As for kinematic distances, indeed, the peculiar motions affect them.  
As shown in Table \ref{tab-kd}, the standard kinematic distance, which do not consider the average peculiar motions of sources 
(\ie\ $\overline{U_s}$ = $\overline{V_s}$ = $\overline{W_s}$ = 0 km s$^{-1}$), 
are biased higher than the true (parallax) values in the second quadrant.  
The ``revised kinematic distances'' \citep{reid09b} using the average peculiar motion of the Perseus arm ($\overline{U_s}$ =   9.2 km s$^{-1}$,
$\overline{V_s}$ = --8.0 km s$^{-1}$, and $\overline{W_s}$ = --2.3 km s$^{-1}$)
have better agreement with the distance from the parallax than the standard kinematic distance in the second quadrant.
In the third quadrants, revised kinematic distances may be somewhat biased toward larger distances compared to parallax values. 
This portion of the Perseus arm may have a smaller $\overline{U_s}$ peculiar motion than the average for all sources. 
Therefore, when calculating (revised) kinematic distances, for Perseus arm sources in the third quadrant, 
one should probably use $\overline{U_s}$ $\sim$ 2 km s$^{-1}$. 
Clearly, more parallax/proper motion measurements are needed to better understand average peculiar motions along the Perseus arm.

\begin{deluxetable}{llrrrl}
\tablecaption{Peculiar Motions of Sources in the Perseus Arm \label{tab-uvw}}
\tablewidth{0pt}
\tablehead{
\colhead{Source} & \colhead{Alias} & \colhead{U$_s$} & \colhead{V$_s$} & \colhead{W$_s$} & \colhead{Ref.} \\
                & & (km s$^{-1}$) & (km s$^{-1}$) & (km s$^{-1}$)  
}
\startdata
G043.16+0.01 & W49(N) & --7.6 $\pm$ \ 11.6 & --17.1 $\pm$ 15.2 & 11.7 $\pm$ 10.4 & 14\\
G048.60+0.02 &                  & --8.0 $\pm$ \ 9.9 & 9.1 $\pm$ 12.5      & 7.1 $\pm$ \ 6.7  & 14 \\
G094.60--1.79 & AFGL 2789 &  4.4 $\pm$ \ 9.7 & --25.6 $\pm$ \ 7.5 & --11.2 $\pm$ 10.3  & 1, 7\\ 
G095.29--0.93 &	 	& --10.3 $\pm$ \ 6.2 & --7.0 $\pm$ \ 5.6 & 3.7 $\pm$ \ 4.7 & 1 \\ 
G100.37--3.57 & 		& 11.7 $\pm$ 10.3 & --3.1 $\pm$ 10.5 & 4.3 $\pm$ \ 9.9  & 1\\
G108.20+0.58  & 		& --12.7 $\pm$ 11.0 & --15.2 $\pm$ 10.3 & 10.7 $\pm$ 11.0 & 1 \\
G108.47--2.81 & 		& 19.1 $\pm$ \ 7.2 & --13.1 $\pm$ \ 6.5 & --7.6 $\pm$ \ 7.2  & 1\\
G108.59+0.49  & 		& 45.1 $\pm$ \ 5.6 & \ --4.6 $\pm$ \ 7.0 & \ 1.1 $\pm$ \ 4.9  & 1\\
G111.23--1.23 & 		& 31.7 $\pm$ 10.7 & 3.4 $\pm$ 25.8 & --0.6 $\pm$ 11.6  & 1\\
G111.25--0.77 & 		& --1.7 $\pm$ \ 8.0 & --11.0 $\pm$ \ 7.2 & --13.3 $\pm$ \ 8.5  & 1\\
G111.54--0.77 & NGC 7538 & 17.1 $\pm$ \ 4.7 & --23.9 $\pm$ \ 5.1 & --9.4 $\pm$ \ 3.3  & 5 \\
G122.01--7.08 & IRAS 00420+5530 & 29.5 $\pm$ \ 5.7 & --6.1 $\pm$ \ 5.7 & 3.9 $\pm$ \ 5.2 & 4 \\
G123.06--6.30 & NGC 281 & 5.4 $\pm$ \ 7.5 & 8.0 $\pm$ \ 7.7 & --15.9 $\pm$ \ 9.9  & 11\\
G123.06--6.30 & NGC 281 (W) & 6.8 $\pm$ \ 4.1 & 4.1 $\pm$ \ 4.0 & --7.3 $\pm$ \ 3.5 & 9 \\
G133.94+1.06 & W3(OH) & 20.2 $\pm$ \ 3.7 & --12.0 $\pm$ \ 3.7 & 2.5 $\pm$ \ 3.1  & 3, 13\\
G134.62--2.19 & S Per & 2.6 $\pm$ \ 5.3 & --13.9 $\pm$ \ 4.8  & --4.4 $\pm$ \ 4.0  & 2 \\
G170.66--0.24 & IRAS 05168+3634 & 11.1 $\pm$ \ 5.6 & --12.7 $\pm$ \ 8.9  & --5.6 $\pm$ \ 9.9 & 10  \\
G183.72--3.66 & & --0.6 $\pm$ \ 5.4 & 1.7 $\pm$ \ 9.3 & 5.0 $\pm$ \ 9.1  & 1\\
G188.79+1.03 &  IRAS 06061+2151& 14.7 $\pm$ \ 5.8 & --20.8 $\pm$ 10.1 & --10.1 $\pm$ \ 7.3 & 6 \\
G188.94+0.88 & S 252   & 0.5 $\pm$ \ 5.3 & --8.0 $\pm$ \ 3.7 & --0.6 $\pm$ \ 3.1  & 8 \\
G192.16--3.81 &     &  3.2 $\pm$ \ 5.3 & --4.8 $\pm$ \ 6.4 & 5.4 $\pm$ \ 5.8  & 12 \\
G192.60--0.04 & S 255 &  1.6 $\pm$ \ 5.7  & 3.5 $\pm$ 12.4 & 4.7 $\pm$ \ 8.1  & 9 \\
G229.57+0.15 &  & 15.1 $\pm$ 12.2 & --12.7 $\pm$ 14.3 & --8.7 $\pm$ 15.3  & 1\\
G236.81+1.98 & & --9.2 $\pm$ \ 7.1 & 1.2 $\pm$ \ 7.6 & --4.4 $\pm$ \ 7.1  & 1\\
G240.31+0.07 &  & --10.5 $\pm$ \ 6.1 & 4.3 $\pm$ \ 8.6 & --12.9 $\pm$ \ 5.5 & 1 \\
\tableline
Average & & 9.2 $\pm$ \ 1.2 & --8.0 $\pm$ \ 1.3 & --2.3 $\pm$ \ 1.1 \\
\enddata
\tablecomments{We assumed 
$R_0$ = 8.33 $\pm$ 0.16 kpc, $\Theta_0$ = 243 $\pm$ 6 km s$^{-1}$, 
dT/dR = --0.2 $\pm$ 0.4 km s$^{-1}$ kpc$^{-1}$  
and Solar Motion components of $U_{\odot}$ = 10.7 $\pm$ 1.8 km s$^{-1}$, 
$V_{\odot}$ = 12.2 $\pm$ 2.0 km s$^{-1}$ and 
$W_{\odot}$ = 8.7 $\pm$ 0.9 km s$^{-1}$ from B1 model in \cite{reid13}
to determine the peculiar motions (U$_s$, V$_s$, W$_s$) in Columns 3--5. 
U$_s$ is the motion toward the Galactic center, 
V$_s$ is in direction of Galactic motion, and W$_s$ is toward the North Galactic Pole (NGP).
Column 5 indicates references. 
(1) This paper; (2) \cite{asaki10}; (3) \cite{hachisuka09}; (4) \cite{Moellenbrock09}; (5) \cite{moscadelli09};
(6) \cite{niinuma11}; (7) \cite{oh10}; (8) \cite{reid09a}; (9) \cite{rygl10}; (10) \cite{sakai12};
(11) \cite{sato08}; (12) \cite{shiozaki11}; (13) \cite{xu06}; (14) \cite{zhang13}}
\end{deluxetable}

\begin{deluxetable}{lccc}
\tablecaption{Parallaxes and Kinematic Distances \label{tab-kd}}
\tablewidth{0pt}
\tablehead{
\colhead{Source}  & \colhead{D$_{\pi}$} & \colhead{D$^{\rm Std}_{\rm k}$} & \colhead{D$^{\rm Rev}_{\rm k}$} \\
                  & (kpc)  & (kpc) & (kpc)   
}
\startdata
W49(N)          & 11.11 $^{+0.79}_{-0.69}$ & 11.61 $^{+0.46}_{-0.45}$ & 10.85 $^{+0.43}_{-0.42}$ \\
G048.60+0.02    & 10.75 $^{+0.61}_{-0.55}$ & 9.95 $^{+0.47}_{-0.47}$ & 9.17 $^{+0.46}_{-0.47}$  \\
G094.60--1.79   & 3.95 $^{+0.41}_{-0.34}$  & 5.65 $^{+0.62}_{-0.62}$ & 4.62 $^{+0.60}_{-0.61}$   \\ 
G095.29--0.93   & 4.85 $^{+0.17}_{-0.16}$  & 4.86 $^{+0.62}_{-0.62}$ & 3.84 $^{+0.61}_{-0.63}$    \\ 
G100.37--3.57   & 3.46 $^{+0.20}_{-0.18}$  & 4.23 $^{+0.61}_{-0.61}$ & 3.23 $^{+0.60}_{-0.61}$    \\
G108.20+0.58    & 4.41 $^{+0.86}_{-0.62}$  & 4.65 $^{+0.63}_{-0.61}$ & 3.62 $^{+0.59}_{-0.58}$    \\
G108.47--2.81   & 3.24 $^{+0.11}_{-0.10}$  & 5.08 $^{+0.64}_{-0.62}$ & 4.04 $^{+0.60}_{-0.59}$    \\
G108.59+0.49    & 2.47 $^{+0.22}_{-0.19}$  & 4.90 $^{+0.64}_{-0.62}$ & 3.85 $^{+0.60}_{-0.58}$    \\
G111.23--1.23   & 3.33 $^{+1.23}_{-0.71}$  & 4.84 $^{+0.65}_{-0.62}$ & 3.79 $^{+0.60}_{-0.58}$    \\
G111.25--0.77   & 3.34 $^{+0.27}_{-0.23}$  & 3.96 $^{+0.61}_{-0.59}$ & 2.96 $^{+0.58}_{-0.57}$    \\
NGC 7538        & 2.65 $^{+0.12}_{-0.11}$  & 5.19 $^{+0.66}_{-0.63}$ & 4.10 $^{+0.61}_{-0.59}$ \\
IRAS 00420+5530 & 2.17 $^{+0.05}_{-0.04}$  & 4.22 $^{+0.67}_{-0.63}$ & 3.18 $^{+0.60}_{-0.58}$ \\
NGC 281         & 2.82 $^{+0.26}_{-0.22}$  & 2.49 $^{+0.58}_{-0.55}$ & 1.57 $^{+0.54}_{-0.52}$ \\
NGC 281(W)      & 2.38 $^{+0.13}_{-0.12}$  & 2.41 $^{+0.57}_{-0.55}$ & 1.50 $^{+0.53}_{-0.52}$ \\
W3(OH)          & 1.95 $^{+0.04}_{-0.03}$  & 4.13 $^{+0.78}_{-0.72}$ & 2.93 $^{+0.68}_{-0.63}$ \\
S Per           & 2.42 $^{+0.11}_{-0.09}$  & 3.34 $^{+0.72}_{-0.66}$ & 2.24 $^{+0.63}_{-0.58}$ \\
IRAS 05168+3634 & 1.88 $^{+0.21}_{-0.17}$  & 8.45 $^{+9.19}_{-4.39}$ & 2.77 $^{+3.54}_{-2.17}$ \\
G183.72--3.66   & 1.59 $^{+0.03}_{-0.03}$  & 1.72 $^{+11.36}_{-1.72}$ & 22.48 $^{+13.98}_{-19.33}$    \\
IRAS 06061+2151 & 2.02 $^{+0.13}_{-0.12}$  & 0.00 $^{+0.00}_{-0.00}$ & 0.76 $^{+2.46}_{-0.76}$ \\
S 252           & 2.10 $^{+0.03}_{-0.03}$  & 2.27 $^{+3.24}_{-2.02}$ & 6.47 $^{+7.55}_{-3.75}$ \\
G192.60--0.04   & 1.52 $^{+0.09}_{-0.09}$  & 0.95 $^{+1.67}_{-0.95}$ & 2.80 $^{+2.58}_{-1.78}$ \\
S 255           & 1.59 $^{+0.07}_{-0.06}$  & 1.13 $^{+1.67}_{-1.13}$ & 3.03 $^{+2.60}_{-1.80}$ \\
G229.57+0.15    & 4.59 $^{+0.27}_{-0.24}$  & 3.97 $^{+0.73}_{-0.67}$ & 4.30 $^{+0.80}_{-0.74}$    \\
G236.81+1.98    & 3.07 $^{+0.27}_{-0.23}$  & 3.56 $^{+0.64}_{-0.60}$ & 3.74 $^{+0.69}_{-0.65}$    \\
G240.31+0.07    & 5.32 $^{+0.49}_{-0.42}$  & 5.92 $^{+0.78}_{-0.72}$ & 6.24 $^{+0.85}_{-0.79}$    \\
\enddata
\tablecomments{Column 1 lists source names. D$_{\pi}$ is the distance converted from the measured parallax.
D$^{\rm Std}_{\rm k}$ is the kinematic distance without considering the average peculiar motions.
D$^{\rm Rev}_{\rm k}$ is the revised kinematic distance using  
$\overline{U_s}$ =   9.2 km s$^{-1}$,
$\overline{V_s}$ = --8.0 km s$^{-1}$, and 
$\overline{W_s}$ = --2.3 km s$^{-1}$. 
The kinematic distances are calculated for 
$R_0$ = 8.33 kpc, $\Theta_0$ = 243 km s$^{-1}$, 
dT/dR = --0.2 km s$^{-1}$ kpc$^{-1}$  
and Solar Motion components of $U_{\odot}$ = 10.7 km s$^{-1}$, 
$V_{\odot}$ = 12.2 km s$^{-1}$ and 
$W_{\odot}$ = 8.7 km s$^{-1}$ from B1 model in \cite{reid13}. 
When we determine the uncertainties in the kinematic distances, we considered a 7 km s$^{-1}$ uncertainty in $V_{\rm LSR}$.   
Kinematic distances for sources near the Galactic anticenter intrinsically have large uncertainties and, 
thus, kinematic distances should be avoided for these sources.}
\end{deluxetable}

\section{CONCLUSION}

We measured parallaxes and proper motions of 12 massive star forming regions 
in the outer portion of the Perseus arm in the second and third Galactic quadrants. 
Combined with 14 results from the literature, we estimated the pitch angle of this
section of the Perseus arm to be 9$^\circ$.9 $\pm$ 1$^\circ$.5. 
We also calculated the three-dimensional Galactic motions and find that
on average the sources in the Perseus arm are
moving toward the Galactic center and slower than the circular Galactic rotation.

\acknowledgments

Financial support by the European Research Council for the ERC Advanced Grant GLOSTAR 
(ERC-2009-AdG, grant agreement No. 247078) is gratefully acknowledged.
This work was supported in part by the Chinese National Science Foundation through grants NSF 11133008 
and the Key Laboratory for Radio Astronomy, Chinese Academy of Sciences.

{\it Facilities:} \facility{VLBA}

\appendix

\section{APPENDIX} \label{appendix}

We present details of parallax and proper motion fits for each source. 
Tables \ref{tbl-G094}--\ref{tbl-G240} summarize the results. 
The uncertainties of parallaxes and proper motions in the tables are the formal fitting uncertainties.

\textbf{G094.60--1.79} 
is also known as AFGL 2789, and its parallax was measured to be 0.326 $\pm$ 0.031 mas, 
corresponding to a distance of 3.07 $\pm$ 0.30 kpc, with VERA \citep{oh10}. 
The proper motions were also obtained to be --2.1 $\pm$ 0.2 mas yr$^{-1}$ and --3.6 $\pm$ 0.5 mas yr$^{-1}$
in right ascension and declination, respectively \citep{oh10}. 
Since the H$_2$O masers were not detected at our last (sixth) epoch in our observations, 
we used only 5 epochs with 7 maser spots and 3 background sources for the fitting.  
When we calculated the peculiar motion in Table \ref{tab-uvw}, 
we used an unweighted average of our measurements and \cite{oh10} for the 
parallax and proper motion, yielding a
parallax of 0.29 $\pm$ 0.03 and a proper motion of --2.3 $\pm$ 0.6 mas yr$^{-1}$ 
in right ascension and --3.8 $\pm$ 0.6 mas yr$^{-1}$ in declination. 
The LSR velocity measured from CO emission is --43 km s$^{-1}$ and --49 km s$^{-1}$ from water maser emission.
We adopted --46.0 $\pm$ 5.0 km s$^{-1}$ for the $V_{\rm LSR}$.   

\begin{table*}
\begin{center}
\caption{Parallax and Proper Motion for G094.60--1.79 \label{tbl-G094}}
\begin{tabular}{cccccc}
\hline \hline 
Maser        & Background & $v_{\rm LSR}$ & Parallax          & $\mu_{\alpha}$  & $\mu_{\delta}$   \\
             & Source     & (km s$^{-1}$) & (mas)             & (mas yr$^{-1}$) & (mas yr$^{-1}$)  \\
\hline
G094.60--1.79 & J2137+5101   & --43.78     & 0.293 $\pm$ 0.055 & --2.86 $\pm$ 0.18 & --4.04 $\pm$ 0.17 \\
             & J2150+5103   & --43.78     & 0.362 $\pm$ 0.015 & --2.88 $\pm$ 0.05 & --3.74 $\pm$ 0.26 \\
             & J2145+5147   & --43.78     & 0.281 $\pm$ 0.058 & --2.97 $\pm$ 0.20 & --4.02 $\pm$ 0.18 \\
             & J2137+5101   & --45.89     & 0.298 $\pm$ 0.026 & --2.54 $\pm$ 0.08 & --3.73 $\pm$ 0.11 \\
             & J2150+5103   & --45.89     & 0.298 $\pm$ 0.004 & --2.75 $\pm$ 0.04 & --3.55 $\pm$ 0.01 \\
             & J2145+5147   & --45.89     & 0.286 $\pm$ 0.033 & --2.65 $\pm$ 0.12 & --3.71 $\pm$ 0.10 \\
             & J2137+5101   & --46.31     & 0.291 $\pm$ 0.037 & --2.49 $\pm$ 0.12 & --3.84 $\pm$ 0.12 \\
             & J2150+5103   & --46.31     & 0.289 $\pm$ 0.011 & --2.71 $\pm$ 0.07 & --3.67 $\pm$ 0.03 \\
             & J2145+5147   & --46.31     & 0.279 $\pm$ 0.034 & --2.60 $\pm$ 0.14 & --3.83 $\pm$ 0.09 \\
             & J2137+5101   & --46.73     & 0.265 $\pm$ 0.028 & --2.56 $\pm$ 0.12 & --4.04 $\pm$ 0.07 \\
             & J2150+5103   & --46.73     & 0.276 $\pm$ 0.022 & --2.75 $\pm$ 0.10 & --3.85 $\pm$ 0.06 \\
             & J2145+5147   & --46.73     & 0.260 $\pm$ 0.025 & --2.66 $\pm$ 0.13 & --4.02 $\pm$ 0.06 \\
             & J2137+5101   & --52.21     & 0.230 $\pm$ 0.021 & --2.19 $\pm$ 0.06 & --4.22 $\pm$ 0.20 \\
             & J2150+5103   & --52.21     & 0.221 $\pm$ 0.038 & --2.43 $\pm$ 0.12 & --4.06 $\pm$ 0.27 \\
             & J2145+5147   & --52.21     & 0.218 $\pm$ 0.008 & --2.30 $\pm$ 0.03 & --4.21 $\pm$ 0.19 \\
             & J2137+5101   & --52.63     & 0.190 $\pm$ 0.018 & --2.29 $\pm$ 0.05 & --4.29 $\pm$ 0.16 \\
             & J2150+5103   & --52.63     & 0.184 $\pm$ 0.045 & --2.52 $\pm$ 0.14 & --4.12 $\pm$ 0.24 \\
             & J2145+5147   & --52.63     & 0.175 $\pm$ 0.009 & --2.41 $\pm$ 0.03 & --4.28 $\pm$ 0.15 \\
             & J2137+5101   & --53.05     & 0.217 $\pm$ 0.021 & --2.24 $\pm$ 0.06 & --4.31 $\pm$ 0.19 \\
             & J2150+5103   & --53.05     & 0.213 $\pm$ 0.055 & --2.47 $\pm$ 0.17 & --4.14 $\pm$ 0.27 \\
             & J2145+5147   & --53.05     & 0.204 $\pm$ 0.028 & --2.36 $\pm$ 0.09 & --4.30 $\pm$ 0.18 \\
\hline
             & Combined fit &             & 0.253 $\pm$ 0.009 &                   &                   \\
             & $<\mu>$      & --48.66     &                   & --2.59 $\pm$ 0.05 & --4.02 $\pm$ 0.04 \\
\hline
\end{tabular}
\end{center}
\end{table*}

\textbf{G095.29--0.93}
is associated with an infrared source, 2MASX J21394111+5120356, 
and its radial velocity is --42.4 km s$^{-1}$ from CS(2--1) emission \citep{bronfman96}, 
--36.5 km s$^{-1}$ from CO emission and --41 km s$^{-1}$ from the water maser spectrum.  
We adopted --38.0 $\pm$ 5.0 km s$^{-1}$ for the $V_{\rm LSR}$.  
  The parallax and proper motions are obtained from 3 maser spots and 4 background sources.

\begin{table*}
\begin{center}
\caption{Parallax and Proper Motion for G095.30--0.93 \label{tbl-G095}}
\begin{tabular}{cccccc}
\hline \hline 
Maser        & Background & $v_{\rm LSR}$ & Parallax          & $\mu_{\alpha}$  & $\mu_{\delta}$   \\
             & Source     & (km s$^{-1}$) & (mas)             & (mas yr$^{-1}$) & (mas yr$^{-1}$)  \\
\hline
G095.29--0.93 & J2137+5101   & --35.25     & 0.231 $\pm$ 0.010 & --2.65 $\pm$ 0.03 & --2.82 $\pm$ 0.07 \\
             & J2150+5103   & --35.25     & 0.223 $\pm$ 0.021 & --2.78 $\pm$ 0.06 & --2.60 $\pm$ 0.11 \\
             & J2145+5147   & --35.25     & 0.209 $\pm$ 0.005 & --2.75 $\pm$ 0.01 & --2.78 $\pm$ 0.07 \\
             & J2139+5300   & --35.25     & 0.202 $\pm$ 0.007 & --2.73 $\pm$ 0.02 & --2.75 $\pm$ 0.08 \\
 & J2137+5101   & --35.67     & 0.216 $\pm$ 0.013 & --2.70 $\pm$ 0.04 & --2.81 $\pm$ 0.04 \\
             & J2150+5103   & --35.67     & 0.214 $\pm$ 0.019 & --2.83 $\pm$ 0.05 & --2.59 $\pm$ 0.09 \\
             & J2145+5147   & --35.67     & 0.200 $\pm$ 0.008 & --2.80 $\pm$ 0.02 & --2.77 $\pm$ 0.05 \\
             & J2139+5300   & --35.67     & 0.190 $\pm$ 0.008 & --2.78 $\pm$ 0.02 & --2.74 $\pm$ 0.04 \\
 & J2137+5101   & --36.10     & 0.211 $\pm$ 0.020 & --2.67 $\pm$ 0.05 & --2.89 $\pm$ 0.09 \\
             & J2150+5103   & --36.10     & 0.210 $\pm$ 0.009 & --2.80 $\pm$ 0.02 & --2.67 $\pm$ 0.12 \\
             & J2145+5147   & --36.10     & 0.191 $\pm$ 0.018 & --2.77 $\pm$ 0.05 & --2.85 $\pm$ 0.08 \\
             & J2139+5300   & --36.10     & 0.180 $\pm$ 0.019 & --2.76 $\pm$ 0.05 & --2.82 $\pm$ 0.08 \\
\hline
             & Combined fit &             & 0.206 $\pm$ 0.004 &                   &                   \\
             & $<\mu>$      & --35.67     &                   & --2.75 $\pm$ 0.01 & --2.76 $\pm$ 0.02 \\
\hline
\end{tabular}
\end{center}
\end{table*}

\textbf{G100.37--3.57}
is associated with the HII region CPM 37 (IRAS22142+5206). 
H$_2$O maser emission was observed between --50 and --15 km s$^{-1}$, 
and it is highly variable over the observation period.
The radial velocity is --36.5 km s$^{-1}$ from CO emission. 
We adopted --37.0 $\pm$ 10.0 km s$^{-1}$ for $V_{\rm LSR}$. 

\begin{table*}
\begin{center}
\caption{Parallax and Proper Motion for G100.37--3.57 \label{tbl-G100}}
\begin{tabular}{cccccc}
\tableline\tableline 
Maser        & Background & $v_{\rm LSR}$ & Parallax          & $\mu_{\alpha}$  & $\mu_{\delta}$   \\
             & Source     & (km s$^{-1}$) & (mas)             & (mas yr$^{-1}$) & (mas yr$^{-1}$)  \\
\tableline
G100.37--3.57 & J2217+5202  & --38.37 & 0.304 $\pm$ 0.016 & --3.73 $\pm$ 0.05 & --2.95 $\pm$ 0.04 \\
                            & J2209+5158 & --38.37 & 0.303 $\pm$ 0.045 & --3.42 $\pm$ 0.15 & --2.90 $\pm$ 0.16 \\
                            & J2209+5158 & --38.37 & 0.310 $\pm$ 0.015 & --3.63 $\pm$ 0.07 & --3.24 $\pm$ 0.04 \\
\tableline                            
                           & Combined fit   & & 0.289 $\pm$ 0.016  \\
                           & $<\mu>$      &   --38.37 &                   & --3.66 $\pm$ 0.05 & --3.02 $\pm$ 0.05 \\
\tableline
\end{tabular}
\end{center}
\end{table*}

\textbf{G108.20+0.58} 
has a radial velocity of --49.2 km s$^{-1}$ from CS(2--1) emission \citep{bronfman96}. 
The parallax is measured using 2 maser spots and 3 background sources. 
We used --49.0 $\pm$ 5.0 km s$^{-1}$ for $V_{\rm LSR}$.

\begin{table*}
\begin{center}
\caption{Parallax and Proper Motion for G108.20+0.58 \label{tbl-G108A}}
\begin{tabular}{cccccc}
\hline \hline 
Maser        & Background & $v_{\rm LSR}$ & Parallax          & $\mu_{\alpha}$  & $\mu_{\delta}$   \\
             & Source     & (km s$^{-1}$) & (mas)             & (mas yr$^{-1}$) & (mas yr$^{-1}$)  \\
\hline
G108.20+0.58 & J2243+6055   & --53.74     & 0.222 $\pm$ 0.046 & --1.87 $\pm$ 0.15 & --1.36 $\pm$ 0.11 \\
             & J2254+6209   & --53.74     & 0.285 $\pm$ 0.049 & --2.33 $\pm$ 0.13 & --1.05 $\pm$ 0.27 \\
             & J2257+5720   & --53.74     & 0.196 $\pm$ 0.054 & --2.52 $\pm$ 0.15 & --0.58 $\pm$ 0.19 \\
 & J2243+6055   & --54.16     & 0.218 $\pm$ 0.038 & --1.88 $\pm$ 0.12 & --1.35 $\pm$ 0.11 \\
             & J2254+6209   & --54.16     & 0.267 $\pm$ 0.057 & --2.34 $\pm$ 0.15 & --1.06 $\pm$ 0.27 \\
             & J2257+5720   & --54.16     & 0.175 $\pm$ 0.050 & --2.54 $\pm$ 0.14 & --0.58 $\pm$ 0.20 \\
\hline 
             & Combined fit &             & 0.227 $\pm$ 0.026 &                   &                   \\
             & $<\mu>$      & --53.95     &                   & --2.25 $\pm$ 0.07 & --1.00 $\pm$ 0.09 \\ 
\hline
\end{tabular}
\end{center}
\end{table*}

\begin{figure*}
\includegraphics[angle=0,scale=0.6]{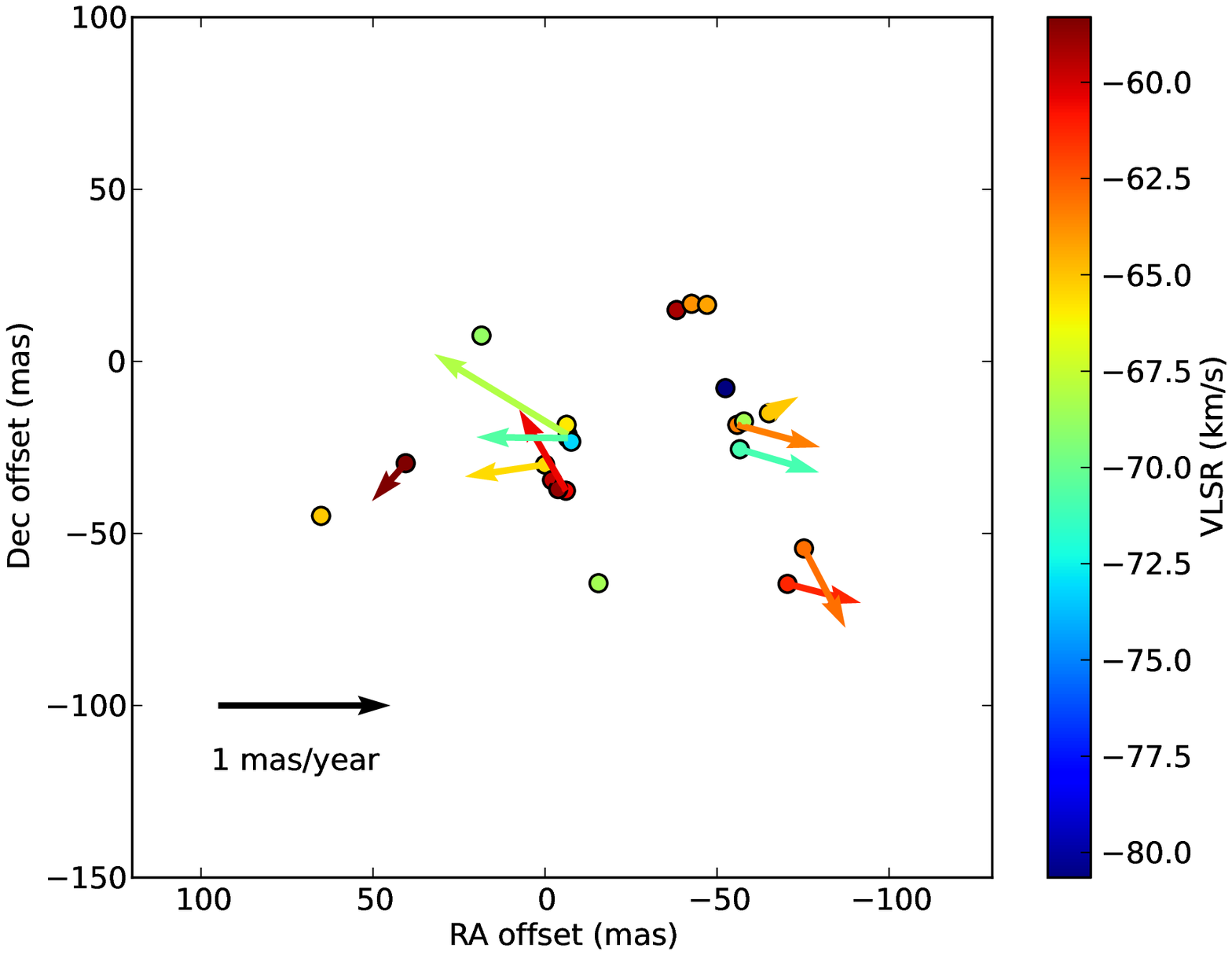}
\caption{Locations and motions of water masers toward G108.47--2.81.  
Motions are with respect to the center of the expansion (0, 0).
Color denotes the LSR velocities of maser spots. 
(A color version of this figure is available in the online journal.)
\label{G108Binternal}}
\end{figure*}

\textbf{G108.47--2.81} 
is associated with IRAS 23004+5642.
We used 6 maser spots and 3 background sources to fit parallax and proper motions.
The radial velocity is --54 km s$^{-1}$ from CO emission, and we used this value to obtain the peculiar motions.
We fitted the data of the relative motions with respect to the reference maser spot ($V_{\rm LSR}$ = --65.48 km s$^{-1}$)
to expansion model \citep{sato10}. 
The expansion velocity is 3.4 $\pm$ 7.4 km s$^{-1}$ and the center of expansion is (0.00 $\pm$ 0.28, 0.03 $\pm$ 0.17) arcsec.
The velocity components are ($V_{0x}$, $V_{0y}$, $V_{0r}$) = (--14.1 $\pm$ 7.6, 2.2 $\pm$ 7.0, --59.3 $\pm$ 3.5) km s$^{-1}$. 
These values correspond to $\mu_x$ = --0.92 $\pm$ 0.49 mas yr$^{-1}$ and $\mu_y$ = 0.14 $\pm$ 0.46 mas yr$^{-1}$ at the distance of 3.24 kpc. Adding these motions to the absolute motion of the reference maser spot, 
we obtained an absolute proper motion of the central star to be 
$\mu_x$ = --3.13 $\pm$ 0.49 mas yr$^{-1}$ and $\mu_y$ = --2.79 $\pm$ 0.46 mas yr$^{-1}$.
Figure \ref{G108Binternal} shows spatial distribution and relative motions with respect to the center of the expansion of water masers toward G108.47--2.81.

\begin{table*}
\begin{center}
\caption{Parallax and Proper Motion for G108.47--2.81 \label{tbl-G108B}}
\begin{tabular}{cccccc}
\tableline\tableline 
Maser        & Background & $v_{\rm LSR}$ & Parallax          & $\mu_{\alpha}$  & $\mu_{\delta}$   \\
             & Source     & (km s$^{-1}$) & (mas)             & (mas yr$^{-1}$) & (mas yr$^{-1}$)  \\
\tableline
G108.47--2.81 & J2301+5706   & --54.52     & 0.310 $\pm$ 0.010 & --3.28 $\pm$ 0.03 & --3.28 $\pm$ 0.06 \\
             & J2258+5719   & --54.52     & 0.306 $\pm$ 0.004 & --3.15 $\pm$ 0.01 & --3.34 $\pm$ 0.03 \\
             & J2257+5720   & --54.52     & 0.328 $\pm$ 0.010 & --3.23 $\pm$ 0.03 & --3.30 $\pm$ 0.05 \\
 & J2301+5706   & --54.52     & 0.286 $\pm$ 0.019 & --1.73 $\pm$ 0.05 & --2.67 $\pm$ 0.09 \\
             & J2258+5719   & --54.52     & 0.283 $\pm$ 0.022 & --1.60 $\pm$ 0.06 & --2.72 $\pm$ 0.11 \\
             & J2257+5720   & --54.52     & 0.304 $\pm$ 0.021 & --1.68 $\pm$ 0.06 & --2.69 $\pm$ 0.11 \\
 & J2301+5706   & --54.94     & 0.300 $\pm$ 0.016 & --3.24 $\pm$ 0.04 & --3.33 $\pm$ 0.06 \\
             & J2258+5719   & --54.94     & 0.296 $\pm$ 0.010 & --3.10 $\pm$ 0.03 & --3.39 $\pm$ 0.03 \\
             & J2257+5720   & --54.94     & 0.316 $\pm$ 0.016 & --3.18 $\pm$ 0.04 & --3.35 $\pm$ 0.06 \\
& J2301+5706   & --54.94     & 0.292 $\pm$ 0.015 & --1.77 $\pm$ 0.04 & --2.66 $\pm$ 0.09 \\
             & J2258+5719   & --54.94     & 0.289 $\pm$ 0.016 & --1.64 $\pm$ 0.04 & --2.72 $\pm$ 0.11 \\
             & J2257+5720   & --54.94     & 0.310 $\pm$ 0.018 & --1.72 $\pm$ 0.05 & --2.68 $\pm$ 0.10 \\
& J2301+5706   & --55.37     & 0.312 $\pm$ 0.008 & --1.79 $\pm$ 0.02 & --2.68 $\pm$ 0.11 \\
             & J2258+5719   & --55.37     & 0.308 $\pm$ 0.010 & --1.66 $\pm$ 0.02 & --2.74 $\pm$ 0.14 \\
             & J2257+5720   & --55.37     & 0.331 $\pm$ 0.010 & --1.74 $\pm$ 0.03 & --2.70 $\pm$ 0.12 \\
 & J2301+5706   & --55.79     & 0.335 $\pm$ 0.006 & --1.80 $\pm$ 0.02 & --2.73 $\pm$ 0.14 \\
             & J2258+5719   & --55.79     & 0.331 $\pm$ 0.006 & --1.67 $\pm$ 0.02 & --2.79 $\pm$ 0.16 \\
             & J2257+5720   & --55.79     & 0.354 $\pm$ 0.007 & --1.75 $\pm$ 0.02 & --2.75 $\pm$ 0.15 \\
\hline
             & Combined fit &             & 0.309 $\pm$ 0.004 &                   &                   \\
             & $<\mu>$      &    --55.01  &                   &  --2.21 $\pm$ 0.08 &  --2.93 $\pm$ 0.04 \\
     
\tableline
\end{tabular}
\end{center}
\end{table*}

\textbf{G108.59+0.49}
 is associated with IRAS 22506+5944 
and its radial velocity is measured to be --52 km s$^{-1}$ from CO emission.
We fitted parallax and proper motions using 2 maser spots and 4 background sources.

\begin{table*}
\begin{center}
\caption{Parallax and Proper Motion for G108.59+0.49 \label{tbl-G108C}}
\begin{tabular}{cccccc}
\hline \hline 
Maser        & Background & $v_{\rm LSR}$ & Parallax          & $\mu_{\alpha}$  & $\mu_{\delta}$   \\
             & Source     & (km s$^{-1}$) & (mas)             & (mas yr$^{-1}$) & (mas yr$^{-1}$)  \\
\hline
G108.59+0.49 & J2243+6055   & --53.47    & 0.273 $\pm$ 0.017 & --5.45 $\pm$ 0.44 & --3.25 $\pm$ 0.03 \\
             & J2254+6209   & --53.47    & 0.420 $\pm$ 0.059 & --5.39 $\pm$ 0.17 & --3.68 $\pm$ 0.20 \\
             & J2301+5706   & --53.47    & 0.336 $\pm$ 0.039 & --5.73 $\pm$ 0.10 & --3.19 $\pm$ 0.26 \\
             & J2258+5719   & --53.47    & 0.315 $\pm$ 0.027 & --5.69 $\pm$ 0.07 & --3.25 $\pm$ 0.31 \\
 & J2243+6055   & --53.89    & 0.494 $\pm$ 0.090 & --5.44 $\pm$ 0.25 & --3.42 $\pm$ 0.32 \\
             & J2254+6209   & --53.89    & 0.449 $\pm$ 0.065 & --5.36 $\pm$ 0.17 & --3.72 $\pm$ 0.30 \\
             & J2301+5706   & --53.89    & 0.354 $\pm$ 0.031 & --5.69 $\pm$ 0.08 & --3.23 $\pm$ 0.33 \\
             & J2258+5719   & --53.89    & 0.335 $\pm$ 0.025 & --5.66 $\pm$ 0.06 & --3.29 $\pm$ 0.35 \\
\hline 
             & Combined fit &            & 0.405 $\pm$ 0.023 &                   &                   \\
             & $<\mu>$      & --53.68    &                   & --5.56 $\pm$ 0.06 & --3.40 $\pm$ 0.08 \\ 
\hline
\end{tabular}
\end{center}
\end{table*}

\textbf{G111.23--1.23} 
is associated with IRAS 23151+5912 
and its radial velocity is --54.4 km s$^{-1}$ from CS(2--1) emission \citep{bronfman96}.
We observed 4 background sources, but J2257+5720 was discarded from the parallax fitting 
since the structure of the quasar changed over time.
We used 2 maser spots and 3 quasars to fit parallax and proper motions.

\begin{table*}
\begin{center}
\caption{Parallax and Proper Motion for G111.23--1.23 \label{tbl-G111A}}
\begin{tabular}{cccccc}
\hline \hline 
Maser        & Background & $v_{\rm LSR}$ & Parallax          & $\mu_{\alpha}$  & $\mu_{\delta}$   \\
             & Source     & (km s$^{-1}$) & (mas)             & (mas yr$^{-1}$) & (mas yr$^{-1}$)  \\
\hline
G111.23--1.23 & J2339+6010   & --49.94     & 0.348 $\pm$ 0.087 & --3.28 $\pm$ 0.48 & --1.80 $\pm$ 0.20 \\
             & J2258+5719   & --49.94     & 0.308 $\pm$ 0.084 & --4.78 $\pm$ 0.36 & --2.80 $\pm$ 0.20 \\
             & J2254+6209   & --49.94     & 0.262 $\pm$ 0.115 & --5.06 $\pm$ 0.46 & --2.80 $\pm$ 0.29 \\
 & J2339+6010   & --50.37     & 0.339 $\pm$ 0.083 & --3.26 $\pm$ 0.48 & --1.64 $\pm$ 0.19 \\
             & J2258+5719   & --50.37     & 0.301 $\pm$ 0.086 & --4.76 $\pm$ 0.36 & --2.63 $\pm$ 0.21 \\
             & J2254+6209   & --50.37     & 0.257 $\pm$ 0.119 & --5.04 $\pm$ 0.47 & --2.64 $\pm$ 0.30 \\
\hline
             & Combined fit &             & 0.300 $\pm$ 0.057 &                   &                   \\
             & $<\mu>$      & --50.16     &                   & --4.37 $\pm$ 0.21 & --2.38 $\pm$ 0.15 \\
\hline 
\hline
\end{tabular}
\end{center}
\end{table*}

\begin{figure*}
\includegraphics[angle=0,scale=0.6]{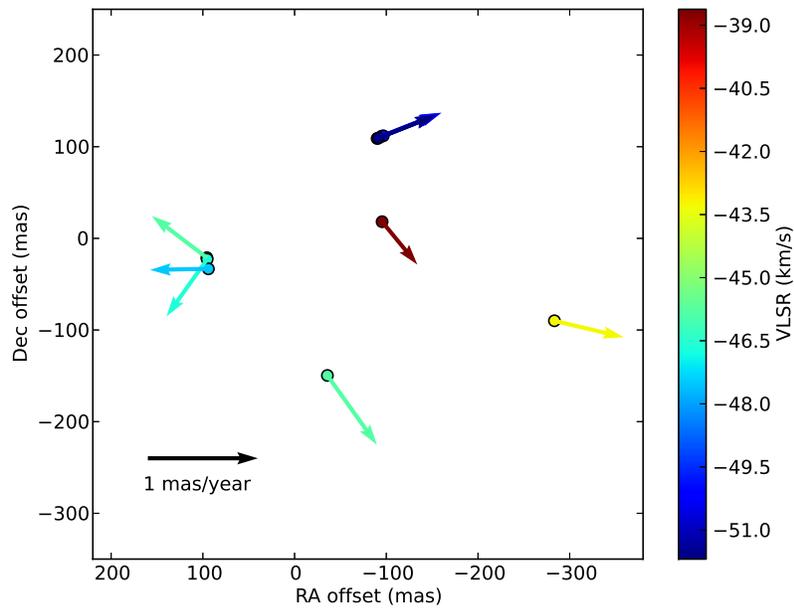}
\caption{
Locations and motions of water masers toward G111.25--0.77.  
Motions are with respect to the center of the expansion (0, 0).
Color denotes the LSR velocities of maser spots. 
(A color version of this figure is available in the online journal.)
\label{G111.25internal}}
\end{figure*}

\textbf{G111.25--0.77}
is associated with IRAS 23139+5939 and its $V_{\rm LSR}$ = --43 km s$^{-1}$ from CO emission. 
We observed 3 background sources. 
However, we did not use J2254+6209 for the parallax fitting,
since it was not detected at the first and sixth epoch.
We obtained parallax and proper motions from 4 maser spots and 2 background sources.
We fitted the data of the relative motions with respect to the reference maser spot ($V_{\rm LSR}$ = --51.26 km s$^{-1}$)
to expansion model \citep{sato10}. 
The expansion velocity is 2.6 $\pm$ 8.1 km s$^{-1}$ and 
the center of expansion is (0.09 $\pm$ 0.50, --0.11 $\pm$ 0.35) arcsec.
The velocity components are ($V_{0x}$, $V_{0y}$, $V_{0r}$) = (8.3 $\pm$ 8.2, --3.3 $\pm$ 8.2, --42.7 $\pm$ 4.0) km s$^{-1}$. 
These values correspond to $\mu_x$ = 0.52 $\pm$ 0.52 mas yr$^{-1}$ and $\mu_y$ = --0.21 $\pm$ 0.52 mas yr$^{-1}$ at the distance of 3.34 kpc. Adding these motions to the absolute motion of the reference maser spot, 
we obtained an absolute proper motion of the central star to be 
$\mu_x$ = --2.02 $\pm$ 0.52 mas yr$^{-1}$ and $\mu_y$ = --2.31 $\pm$ 0.52 mas yr$^{-1}$.
Figure \ref{G111.25internal} shows spatial distribution and relative motions with respect to the center of the expansion of water masers toward G111.25--0.77.

\begin{table*}
\begin{center}
\caption{Parallax and Proper Motion for G111.25--0.77 \label{tbl-G111B}}
\begin{tabular}{cccccc}
\hline \hline 
Maser        & Background & $v_{\rm LSR}$ & Parallax          & $\mu_{\alpha}$  & $\mu_{\delta}$   \\
             & Source     & (km s$^{-1}$) & (mas)             & (mas yr$^{-1}$) & (mas yr$^{-1}$)  \\
\hline
G111.25--0.77 & J2339+6010   & --39.05     & 0.303 $\pm$ 0.018 & --3.09 $\pm$ 0.04 & --1.96 $\pm$ 0.13 \\
             & J2258+5719   & --39.05     & 0.326 $\pm$ 0.024 & --2.81 $\pm$ 0.07 & --2.44 $\pm$ 0.06 \\
 & J2339+6010   & --46.63     & 0.293 $\pm$ 0.018 & --3.22 $\pm$ 0.05 & --2.27 $\pm$ 0.10 \\
             & J2258+5719   & --46.63     & 0.310 $\pm$ 0.024 & --2.94 $\pm$ 0.06 & --2.76 $\pm$ 0.11 \\
 & J2339+6010   & --47.89     & 0.287 $\pm$ 0.021 & --2.17 $\pm$ 0.06 & --1.61 $\pm$ 0.13 \\
             & J2258+5719   & --47.89     & 0.309 $\pm$ 0.022 & --1.89 $\pm$ 0.07 & --2.09 $\pm$ 0.06 \\
 & J2339+6010   & --48.31     & 0.287 $\pm$ 0.022 & --2.23 $\pm$ 0.06 & --1.57 $\pm$ 0.11 \\
             & J2258+5719   & --48.31     & 0.309 $\pm$ 0.020 & --1.94 $\pm$ 0.06 & --2.06 $\pm$ 0.08 \\
\hline 
             & Combined fit &             & 0.299 $\pm$ 0.011 &                   &                   \\
             & $<\mu>$      & --45.47     &                   & --2.54 $\pm$ 0.08 & --2.10 $\pm$ 0.07 \\ 
\hline
\end{tabular}
\end{center}
\end{table*}

\textbf{G183.72--3.66} has a parallax of 0.629 $\pm$ 0.012 mas
based on measurements of two background sources.
The $V_{\rm LSR}$ from CO emission is 2.5 km s$^{-1}$. 
The H$_2$O masers are variable over our observations. 
Since most of maser spots disappeared at the 3rd epoch and only 2 spots survived for more than 2 epochs,
we could not get internal motions.

\begin{table*}
\begin{center}
\caption{Parallax and Proper Motion for G183.72--3.66 \label{tbl-G183}}
\begin{tabular}{cccccc}
\tableline\tableline 
Maser        & Background & $v_{\rm LSR}$ & Parallax          & $\mu_{\alpha}$  & $\mu_{\delta}$   \\
             & Source     & (km s$^{-1}$) & (mas)             & (mas yr$^{-1}$) & (mas yr$^{-1}$)  \\
\tableline
G183.72--3.66 & J0540+2507 & \ \ 4.58       & 0.632 $\pm$ 0.020 & 0.43 $\pm$ 0.05 & --1.32 $\pm$ 0.20 \\
             & J0550+2326 & \ \ 4.58       & 0.627 $\pm$ 0.004 & 0.32 $\pm$ 0.01 & --1.32 $\pm$ 0.08 \\
\tableline             
             &          Combined fit  & & 0.629 $\pm$ 0.012 &  \\
             & $<\mu>$      & 4.58     &                   &  0.38 $\pm$ 0.03 & --1.32 $\pm$ 0.10 \\

\tableline
\end{tabular}
\end{center}
\end{table*}

\textbf{G229.57+0.15}
is associated with IRAS 07207--1435.
The H$_2$O maser emission ranges from 45 km s$^{-1}$ to 60 km s$^{-1}$ with a 
predominantly double-peak spectrum. The radial velocity is 43 km s$^{-1}$ from CO emission. 
We adopted 47 $\pm$ 10 km s$^{-1}$ for $V_{\rm LSR}$.
We measured parallax and proper motions from one maser spot detected at all epochs and 4 background sources. 
Because of the source's low declination, the error in right ascension is much smaller 
than that in declination.  

\begin{table*}
\begin{center}
\caption{Parallax and Proper Motion for G229.57+0.15 \label{tbl-14}}
\begin{tabular}{cccccc}
\tableline\tableline 
Maser        & Background & $v_{\rm LSR}$ & Parallax          & $\mu_{\alpha}$  & $\mu_{\delta}$   \\
             & Source     & (km s$^{-1}$) & (mas)             & (mas yr$^{-1}$) & (mas yr$^{-1}$)  \\
\tableline
G229.57+0.15 & J0721--1530 & \ 57.11        & 0.214 $\pm$ 0.012 & --1.35 $\pm$ 0.03 & 0.74 $\pm$ 0.13 \\
             & J0724--1545 & \ 57.11        & 0.228 $\pm$ 0.015 & --1.25 $\pm$ 0.04 & 0.62 $\pm$ 0.12 \\
             & J0729--1320 & \ 57.11        & 0.193 $\pm$ 0.017 & --1.49 $\pm$ 0.05 & 1.19 $\pm$ 0.15 \\
             & J0721--1630 & \ 57.11        & 0.239 $\pm$ 0.017 & --1.21 $\pm$ 0.05 & 0.55 $\pm$ 0.28 \\
\tableline             
             &        Combined fit  &  & 0.218 $\pm$ 0.012 &  \\
                          & $<\mu>$      & 57.11     &    &   --1.33 $\pm$ 0.03 & 0.77 $\pm$ 0.10             \\
\tableline
\end{tabular}
\end{center}
\end{table*}

\textbf{G236.81+1.98}
is associated with IRAS 07422--2001 and its radial velocity is 43 km s$^{-1}$ from CO emission.
Four background sources were observed, but two of them (J0735--1735 and J0739--2301) 
were not used for the parallax fitting.
J0735--1735 and J0739--2301 are separated by about 3 degrees from the maser source and mostly in the north-south direction. 
J0735--1735 has structure, which is likely caused by atmospheric distortion 
and J0739--2301 was not detected after the third epoch.  
Because of the source's low declination, the error in right ascension is much smaller 
than that in declination.  

\begin{figure*}
\includegraphics[angle=0,scale=0.6]{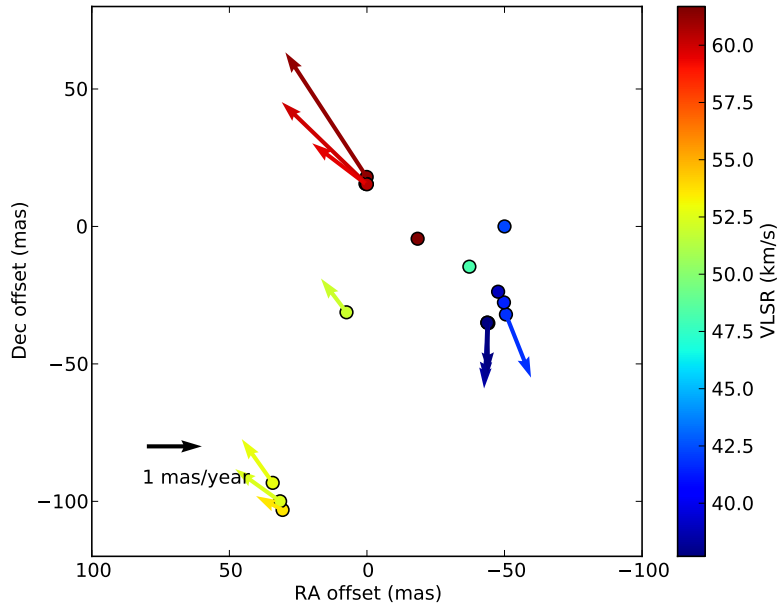}
\caption{ 
Locations and motions of water masers toward G236.81+1.98.  
Motions are with respect to the center of the expansion (0, 0).
Color denotes the LSR velocities of maser spots.   
(A color version of this figure is available in the online journal.)
\label{G236internal}}
\end{figure*}

We fitted the data of the relative motions with respect to the reference maser spot ($V_{\rm LSR}$ = 42.31 km s$^{-1}$)
to expansion model \citep{sato10}. 
The expansion velocity is 0.5 $\pm$ 10.1 km s$^{-1}$ and 
the center of expansion is ( 0.05 $\pm$ 0.76, 0.03 $\pm$ 0.85) arcsec.
The velocity components are ($V_{0x}$, $V_{0y}$, $V_{0r}$) = (10.5 $\pm$ 6.3, 8.2 $\pm$ 6.0, 48.5 $\pm$ 5.9) km s$^{-1}$. 
These values correspond to $\mu_x$ = 0.72 $\pm$ 0.43 mas yr$^{-1}$ and $\mu_y$ = 0.56 $\pm$ 0.41 mas yr$^{-1}$ at the distance of 3.07 kpc. Adding these motions to the absolute motion of the reference maser spot, 
we obtained an absolute proper motion of the central star to be 
$\mu_x$ = --2.49 $\pm$ 0.43 mas yr$^{-1}$ and $\mu_y$ = 2.67 $\pm$ 0.41 mas yr$^{-1}$.
Fig. \ref{G236internal} shows spatial distribution and relative motions with respect to the center of the expansion of water masers toward G236.81+1.98.

\begin{table*}
\begin{center}
\caption{Parallax and Proper Motion for G236.81+1.98 \label{tbl-G236}}
\begin{tabular}{cccccc}
\tableline\tableline 
Maser        & Background & $v_{\rm LSR}$ & Parallax          & $\mu_{\alpha}$  & $\mu_{\delta}$   \\
             & Source     & (km s$^{-1}$) & (mas)             & (mas yr$^{-1}$) & (mas yr$^{-1}$)  \\
\tableline
G236.81+1.98 & J0741--1937 & \ 42.31        & 0.310 $\pm$ 0.018 & --3.14 $\pm$ 0.05 & 1.87 $\pm$ 0.36 \\
             & J0745--1828 & \ 42.31        & 0.335 $\pm$ 0.024 & --3.29 $\pm$ 0.13 & 2.35 $\pm$ 0.03 \\
       \tableline      
                & Combined fit &   & 0.326 $\pm$ 0.026 & \\
                          & $<\mu>$      & 42.31    &                   & --3.21 $\pm$ 0.07 & 2.11 $\pm$ 0.18  \\
\tableline
\end{tabular}
\end{center}
\end{table*}

\textbf{G240.31+0.07}
is associated with 2MASS J07445196--2407399 and its $V_{\rm LSR}$ is 67.0 km s$^{-1}$ 
obtained from CO emission. 
The parallax was measured using 3 maser spots and 3 background sources.
Because of the source's low declination, the error in right ascension is much smaller 
than that in declination.   

\begin{table*}
\begin{center}
\caption{Parallax and Proper Motion for G240.31+0.07 \label{tbl-G240}}
\begin{tabular}{cccccc}
\tableline\tableline 
Maser        & Background & $v_{\rm LSR}$ & Parallax          & $\mu_{\alpha}$  & $\mu_{\delta}$   \\
             & Source     & (km s$^{-1}$) & (mas)             & (mas yr$^{-1}$) & (mas yr$^{-1}$)  \\
\tableline
G240.31+0.07 & J0745--2451 & \ 67.95        & 0.194 $\pm$ 0.019 & --2.38 $\pm$ 0.05 & 2.47 $\pm$ 0.24 \\
             & J0749--2344 & \ 67.95        & 0.152 $\pm$ 0.018 & --2.65 $\pm$ 0.05 & 3.13 $\pm$ 0.14 \\
             & J0740--2444 & \ 67.95        & 0.227 $\pm$ 0.066 & --2.22 $\pm$ 0.21 & 2.14 $\pm$ 0.30 \\
 & J0745--2451 & \ 67.53        & 0.195 $\pm$ 0.018 & --2.36 $\pm$ 0.05 & 2.08 $\pm$ 0.14 \\
             & J0749--2344 & \ 67.53        & 0.210 $\pm$ 0.017 & --2.47 $\pm$ 0.04 & 2.95 $\pm$ 0.12 \\
             & J0740--2444 & \ 67.53        & 0.183 $\pm$ 0.021 & --2.35 $\pm$ 0.06 & 2.19 $\pm$ 0.16 \\
 & J0745--2451 & \ 67.11        & 0.190 $\pm$ 0.016 & --2.42 $\pm$ 0.04 & 2.16 $\pm$ 0.16 \\
             & J0749--2344 & \ 67.11        & 0.180 $\pm$ 0.017 & --2.58 $\pm$ 0.05 & 3.15 $\pm$ 0.13 \\
             & J0740--2444 & \ 67.11        & 0.218 $\pm$ 0.022 & --2.27 $\pm$ 0.06 & 1.90 $\pm$ 0.27 \\                          
\tableline             
             &           Combined fit  & & 0.188 $\pm$ 0.009 & \\
             & $<\mu>$      & 67.53     &                   & --2.43 $\pm$ 0.02 & 2.49 $\pm$ 0.09 \\ 
\tableline
\end{tabular}
\end{center}
\end{table*}

\end{document}